\documentclass[fleqn,usenatbib]{mnras}
\usepackage{graphicx} 
\usepackage{acronym}
\usepackage{braket}
\usepackage{amsmath}
\usepackage{xcolor} 

\title[Prior convictions]{Prior Convictions: Considerations for analysing weak gravitational-wave signals}
\author[D. Williams \& J. Veitch]{Daniel Williams\thanks{daniel.williams@glasgow.ac.uk}$^1$, John Veitch$^1$\\$^1$ Institute for Gravitational Research, School of Physics \& Astronomy, University of Glasgow. G12 8QQ. United Kingdom.}
\date{\today}

\usepackage{color}
\newcommand{\Mc}{{\ensuremath{\mathcal{M}}}}
\newcommand{\rhomf}{{\ensuremath{\rho_\mathrm{mf}}}}
\newcommand{\dl}{{\ensuremath{D_L}}}
\newcommand{\pastro}{\ensuremath{p_{\mathrm{astro}}}}

\definecolor{claudeorange}{rgb}{0.85,0.42,0.05}
\newcommand{\claude}[1]{{\color{claudeorange} #1}}

\acrodef{GW}[GW]{Gravitational wave}
\acrodef{CBC}[CBC]{compact binary coalescence}
\acrodef{LIGO}[LIGO]{the Laser Interferometer Gravitational-wave Observatory}
\acrodef{SNR}[SNR]{signal-to-noise ratio}
\acrodef{JSD}[JSD]{Jensen-Shannon Divergence}
\acrodef{KL}[KL]{Kullback--Leibler}
\acrodef{BBH}[BBH]{binary black hole}
\acrodef{NSBH}[NSBH]{neutron star--black hole binary}
\acrodef{BNS}[BNS]{binary neutron star}
\acrodef{PE}[PE]{parameter estimation}
\acrodef{FAR}[FAR]{false alarm rate}
\acrodef{PSD}[PSD]{power spectral density}
\acrodef{GWTC}[GWTC]{the Gravitational-wave Transient Catalog}
\acrodef{GWTC3}[GWTC--3]{\ac{GWTC}, version 3}
\acrodef{GWTC4}[GWTC--4.0]{\ac{GWTC}, version 4.0}
\acrodef{GWTC5}[GWTC--5.0]{\ac{GWTC}, version 5.0}
\acrodef{ESS}[ESS]{effective sample size}
\newacro{O4}[O4]{the fourth observing run}
\newacro{O4a}[O4a]{the first part of the fourth observing run}
\newacro{O4b}[O4b]{the second part of the fourth observing run}
\newacro{O4c}[O4c]{the third part of the fourth observing run}
\newacro{O3}[O3]{the third observing run}
\newacro{LVK}[LVK]{LIGO--Virgo--KAGRA Collaboration}
\newacro{GWOSC}[GWOSC]{gravitational-wave open science centre}
\newacro{STFC}[STFC]{Science \& Technology Facilities Council}
\defcitealias{wwwb-subthreshold}{Williams et al., in prep.}
\begin{document}

\maketitle

\begin{abstract}
    Analysing quiet gravitational-wave signals often returns a luminosity-distance posterior which is mostly prior. 
    For a prior uniform in Euclidean volume, the posterior has a finite mode only if the matched-filter SNR is at least $\sqrt{8}$; below this the median and credible interval reflect the prior bounds rather than the data. 
    Cosmologically-appropriate priors lower the threshold but do not remove it, and the Malmquist prior, which weights by the probability of detection, counts the data twice. 
    We show that the Jeffreys prior on luminosity distance, conditioned on all other signal parameters, has no such threshold, does not double-count the data, and lets the distance be marginalised analytically and reconstructed.  Reweighting its samples allows the  astrophysical posterior to be recovered.
    We also demonstrate the use of the prior on real events: six near-threshold binary black hole candidates from O3 and the subthreshold candidate GW190531.
\end{abstract}

\section{Introduction}\label{sec:intro}

\ac{GW} signals detected by \ac{LIGO}~\citep{ligo_scientific_collaboration_advanced_2015}, Virgo~\citep{acernese_advanced_2015}, and KAGRA~\citep{2021PTEP.2021eA101A} are commonly analysed in a noise-dominated regime (see \citet{abbott_guide_2020} for a detailed discussion of the sources of this noise). 
Technological improvements over the first decade of detection have markedly improved the sensitivity of these instruments by reducing and eliminating sources of noise.
However, detection candidates are still typically identified low\acp{SNR}: around eight and above~\footnote{The typical \ac{SNR} used in \ac{GW} astronomy is the network \ac{SNR} which is calculated as the sum in quadrature of the \ac{SNR} of the signal in each observatory which is actively observing at the time of the detection.}.
We know that signals from more distant sources are also present in the data, albeit at lower \acp{SNR} than those which are typically regarded as candidates in published catalogues (e.g.~\citep{catalog-gwtc-5}).
Therefore there is an incentive to include these distant but quiet candidates in analyses, since they provide information from the Universe at an earlier stage in its development than louder, closer candidates.
The rate of \ac{BBH} events is expected to vary with redshift, in a way consistent with the evolution of the star formation rate~\citep{pop-gwtc-5d0,madau_dickinson_2014}, and so the inclusion of distant events is important for understanding the population of \ac{GW} sources.

In this paper we show that for such signals the inferred posterior on the luminosity distance can dominated by the prior.
In section~\ref{sec:distance-priors} we show how, for a prior uniform in Euclidean volume, there will be no finite posterior mode below a matched-filter \ac{SNR} of $\rho_{\rm mf} = \sqrt{8}$, and that summary statistics of the resulting long tail will depend on the chosen prior bounds.
We show similar behaviour for other priors commonly used on the luminosity distance.
In section~\ref{sec:jeffreys} we introduce the conditional Jeffreys prior $\propto \rho/\dl$ over the luminosity distance, as a potential solution to this problem.
We also show how this enables analytic marginalisation over the luminosity distance during \ac{PE}; in section~\ref{sec:reweighting} we discuss the possibility of reweighting this prior to more conventional choices of distance prior.

\subsection{\ac{GW} source parameter estimation} \label{sec:parameter-estimation}

The parameters of observed \ac{GW} signals are typically inferred using Bayesian methodologies.
Bayesian inference provides a powerful framework for these analyses, as \ac{GW} events cannot be repeated, and (at least in current detectors) signals are detected in a noise-dominated, low \ac{SNR} regime.
Analyses are typically performed using either stochastic sampling of the posterior~\citep[e.g.][]{code-bilby,code-lalinference}, or by training a machine-learning algorithm to approximate the posterior distribution~\citep[e.g.][]{code-dingo, code-vitamin}.
All of the catalogues which provide parameter estimation results for observed events have used Bayesian techniques~\citep{catalog-4ogc, abbott_gwtc-3_2023, catalog-beyond}.

In the ten years since the first detection of gravitational-waves~\citep{special-gw150914-detection} the routine and frequent detection of signals has become a reality, with hundreds of events identified throughout the literature, including by the LVK Collaboration~\citep[e.g.][for their recent results]{catalog-gwtc-5} and other members of the community (see~\citet{catalog-beyond} for a survey of these prior to the fourth observing run).
Improvements in techniques for identifying events has led to increased confidence in the veracity of these signals.
This has inevitably led to parameter estimation being attempted on signals which have low SNRs~\citep{Huang:2020ysn}.
The results of inference on these signals provide interesting problems with interpretation, as the amount of information conveyed by the data may not be more informative than the prior assumptions throughout the parameter space of the analysis.

\subsection{The problem of prior-dominated results}
\label{sec:problem}

Bayesian parameter estimation aims to determine the probability distribution on a set of unknown model parameters $\theta$ by updating prior knowledge about those parameters, denoted $p(\theta | I)$ with information from data $d$, with other information about the system encapsulated within $I$.
The posterior probability distribution, $p(\theta | d, I)$, is related to the prior and the likelihood $p(d| \theta, I)$ via Bayes' Theorem,
\begin{equation}
    p(\theta | d, I) \propto p(\theta|I) p(d|\theta, I),
\end{equation}
and so is always a combination of the prior and the likelihood.

Depending on the information content of the data, the results of any Bayesian analysis will exist somewhere on a continuum between being exclusively informed by the prior (in the case that the data are uninformative), or exclusively informed by the likelihood (in the case that the prior is uninformative).
Situations at either end of this range are straightforward to interpret, however results in the middle-ground can be less clear. 
Several studies have looked at the effect on the posterior of changing prior assumptions in such cases, particularly on the mass and spin parameters (see e.g. \citep{Vitale:2017cfs,Zevin:2020gxf,Zhu:2020zij,Moore:2021xhn,Prasad:2026idz}).

Analyses of \ac{CBC} sources often choose a prior that is deliberately uninformative on parameters of astrophysical interest (e.g. uniform-in-component masses) so that the marginal posterior distributions on these parameters, and the associated credible intervals, can be said to represent fairly the results from the data alone
For other parameters, the choice of prior is motivated by considering the symmetry of the source, so that the binary has no preferred orbital orientation or phase, which is uninformative when considering the astrophysical population of sources (although not when considering detected population). 
In these cases, any structure in the joint posterior of the orientation parameters can be unambiguously said to be determined by the likelihood.

In the case of the source luminosity distance $\dl$, we are in an intermediate situation where the prior determined only by a naive uniform distribution of sources throughout physical spacetime is informative on the $\dl$ parameter. 
At low redshifts $z\ll 1$ we may approximate the universe as being Euclidean and so
\begin{equation}
p(\dl | E)\propto \frac{\partial V}{\partial \dl}\propto\dl^2.\label{eq:dl^2}
\end{equation}
This distribution is only properly normalised when imposing an upper cutoff in distance, otherwise the prior will grow without limit and dominate any likelihood function.
Thankfully, the observable universe has a finite volume\footnote{For the Planck 2015 cosmology the prior in $\dl$ which is uniform in comoving volume and time peaks at $\dl\approx7.8$\,Gpc ($z\approx1.1$) and decreases beyond it.} and our detectors have limited sensitivity, so we are not confronted with this issue at a fundamental level, but nonetheless results produced even when using a realistic cosmological distance prior usually include an upper limit on the distance that is unphysical, so as to prevent the likelihood from being dominated by the distance prior. 
This works well when the signal analysed is loud, and distance is well-constrained for any reasonable upper limit on the prior, but for weak signals the posterior often contains a long tail toward high distance as a result of the non-uniform prior. 
If the likelihood cannot overcome the distance prior, then the posterior even on normally well-determined parameters such as the chirp mass $\Mc$ often contains a small peak superimposed on a broad, uninformative prior. 
This makes the credible intervals on \emph{any} parameters highly prior-dependent, and intuitively less useful for interpretation of the data.

\begin{figure}
    \centering
    \includegraphics[width=\linewidth]{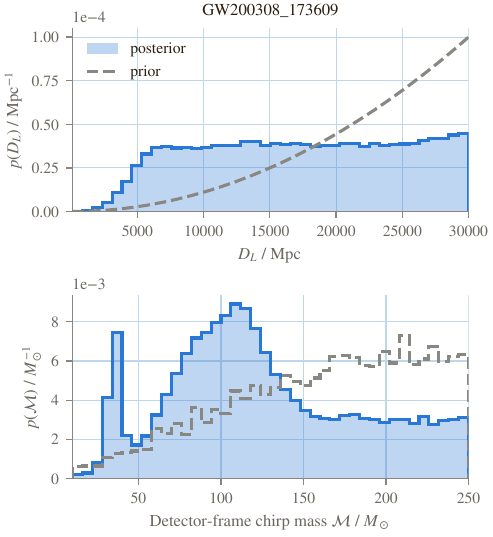}
    \caption{The marginal posterior distribution (filled histogram) of the luminosity distance (top) and detector-frame chirp mass (bottom) for GW200308\_173609, from the \ac{GWTC3} samples released without reweighting to a cosmological prior. The prior is shown as a line (top, $\propto\dl^2$ over $100$--$30\,000$\,Mpc) and as an outline histogram of prior samples (bottom).}
    \label{fig:gw200308}
\end{figure}

\begin{figure}
    \centering
    \includegraphics[width=\linewidth]{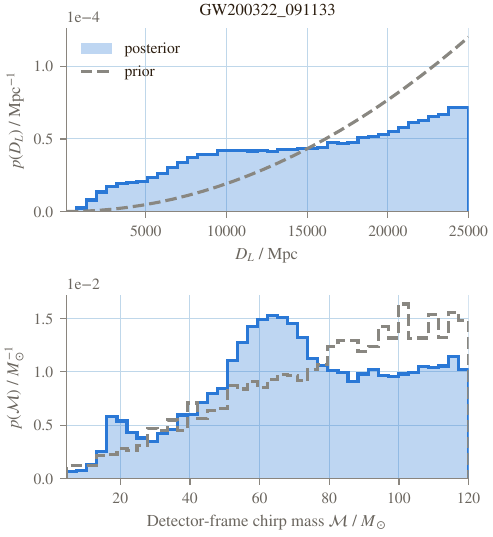}
    \caption{As figure~\ref{fig:gw200308}, for GW200322\_091133. The prior on the luminosity distance is $\propto\dl^2$ over $100$--$25\,000$\,Mpc.}
    \label{fig:gw2003222}
\end{figure}

Two \ac{GW} events which fall into this category are GW200308\_173609 and GW200322\_091133, whose $\dl$ and $\Mc$ marginal posteriors are reproduced using samples from~\citet{ligo_scientific_collaboration_and_virgo_2026_22685054}, in figures \ref{fig:gw200308} and \ref{fig:gw2003222} respectively. 
These were both observed during the third observing run, and appear for the first time in \ac{GWTC3} as relatively low-significance events~\citep{abbott_gwtc-3_2023}.
The search pipelines give GW200308\_173609 a \ac{FAR} between $2.4$ and $69\,000\,\mathrm{yr}^{-1}$ and a network matched-filter \ac{SNR} of $7.9$--$8.3$.
For GW200322\_091133 a \ac{FAR} between $140$ and $1400\,\mathrm{yr}^{-1}$ and an \ac{SNR} of $7.7$--$9.0$.
In both cases the posterior produced during \ac{PE} contained ``low-likelihood'' modes at large distances and high masses.

\begin{figure*}
    \centering
    \includegraphics[width=\textwidth]{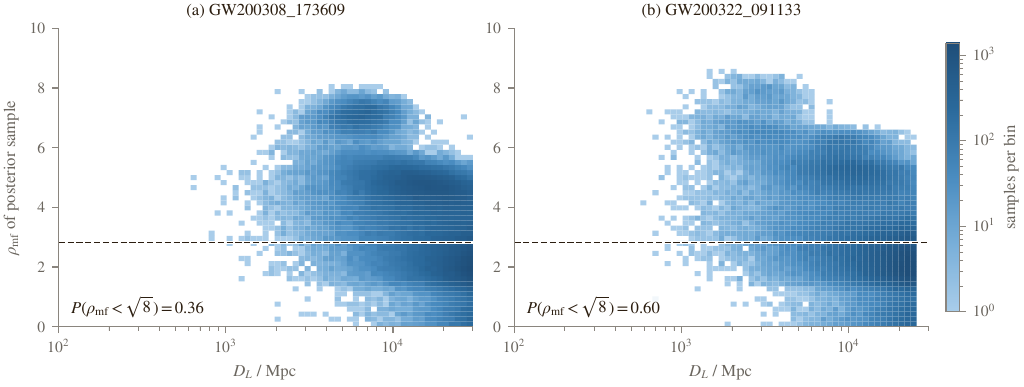}
    \caption{
        The network matched-filter \ac{SNR} of each posterior sample against its luminosity distance for (a) GW200308\_173609 and (b) GW200322\_091133, from the GWTC-3 samples released without reweighting to a cosmological prior, whose prior on distance is $\propto\dl^2$. The dashed line is $\rhomf=\sqrt{8}$, and the text gives the fraction of samples below it.}
    \label{fig:sqrt8-events}
\end{figure*}

We can understand when this behaviour will occur by considering the expansion of the log-posterior in terms of the gravitational wave produced by a source at 1 Mpc such that $h(\dl) = h_1 \times (1\,\mathrm{Mpc}/\dl)$ and using the standard noise-weighted inner product~\citep{Cutler:1994ys},
\begin{equation}
    \langle a | b \rangle = 4 \, \mathrm{Re} \int_0^{\infty} \frac{\tilde{a}^*(f)\, \tilde{b}(f)}{S_n(f)} \, \mathrm{d}f.
\end{equation}
The log-likelihood function up to a constant $c$ is
\begin{eqnarray}\label{eq:loglikelihood}
\ln L +c &= -\frac{1}{2\dl^2}\braket{h_1|h_1} + \frac{1}{\dl}\braket{d|h_1} \nonumber \\
&= -\frac{\rho_1^2}{2\dl^2} + \frac{\rhomf\rho_1}{\dl}
\end{eqnarray}
where $\rhomf = \braket{d|h}/\sqrt{\braket{h|h}}$ is the matched filter \ac{SNR} and $\rho_1=\sqrt{\braket{h_1|h_1}}$ is the optimal \ac{SNR} of the source were it at a fiducial distance of 1\,Mpc. 
The luminosity distance that maximises the likelihood is $\dl^\mathrm{ML} = \rho_1 / \rhomf$.
However if we compare to the value that maximises the posterior, with the Euclidean prior defined in equation~\ref{eq:dl^2}, we find instead that the local maximum of the posterior (the MAP value) is given by
\begin{eqnarray}
\dl^\mathrm{MAP} = \frac{2\rho_1}{\rhomf}\frac{1}{1+\sqrt{1-8/\rhomf^2}},
\end{eqnarray}
This allows us to identify a threshold \ac{SNR} value $\rhomf=\sqrt{8}$, above which the posterior contains a well-defined peak at finite $\dl$, in addition to the prior-domination when $\dl\rightarrow\infty$. Below this $\rhomf$ cutoff however the posterior monotonically increases at higher $\dl$ and we can say that the likelihood is only weakly informative.
Figure~\ref{fig:sqrt8-events} shows the matched-filter \ac{SNR} of the posterior samples released for the two events above: a large fraction of the posterior lies at $\rhomf<\sqrt{8}$ and large distance.

For a \ac{CBC} signal where the prior is allowed to vary over the system's mass and time of coalescence, we typically have a well-localized posterior peak for these parameters, with a high $\rhomf$, surrounded by regions of $\rhomf < \sqrt{8}$ with posteriors that diverge at high distance. 
The relative contribution of the high-\ac{SNR} peak and the low-\ac{SNR} ``sea'' surrounding it will depend on the bounds chosen for the prior, particularly in the $\dl$, $\Mc$ and $t_c$ parameters.
Additionally, when using a nested sampling algorithm to perform parameter estimation with $N_{\rm live}$ live points, the length of the run $N_{\rm iter}$ scales proportionally to the \ac{KL} divergence between the prior and posterior, which is approximately given in nats by the logarithm of the relative volume of the posterior, such that $N_{\rm iter}\approx N_{\rm live}\ln(V_{\rm prior}/V_{\rm post})$~\citep{Skilling:2006gxv}. 
Choosing a very uninformative prior with large bounds also results in a waste of computing time spent sampling large regions of low likelihood.

\begin{figure*}
    \centering
    \includegraphics[width=\textwidth]{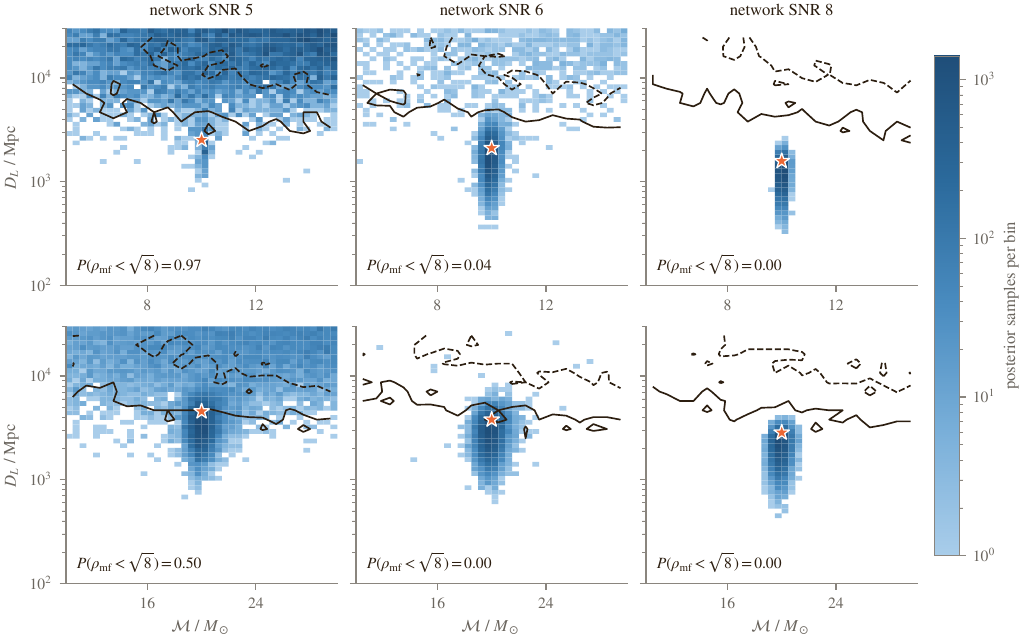}
    \caption{Posterior samples of the luminosity distance and detector-frame chirp mass for injected signals with a network \ac{SNR} of five, six and eight (columns) and chirp masses of $10$ and $20\,M_\odot$ (rows), analysed with the astrophysical distance prior. The white contours enclose $50\%$ (solid) and $90\%$ (dashed) of the prior samples, the star marks the injected values, and the text gives the fraction of posterior samples with a network matched-filter \ac{SNR} below $\sqrt{8}$. Each panel is a single noise realisation, and the chirp-mass prior spans $0.5$--$1.5$ times the injected value.}
    \label{fig:injections}
\end{figure*}

The same behaviour is seen for simulated signals injected into noise.
In figure~\ref{fig:injections} we show how the posteriors inferred from analysing six such signals, at a variety of different \acp{SNR} and chirp masses, vary.
For a signal injected with a network \ac{SNR} of five, the posteriors for chirp masses of $10$ and $20\,M_\odot$ spread over the prior at large distance, with $0.97$ and $0.50$ of the samples at $\rhomf<\sqrt{8}$; for the $10\,M_\odot$ signal, injected at $2.5$\,Gpc, the median posterior distance is $15.6$\,Gpc.
At an \ac{SNR} of six the posteriors are generally confined in a region close to the injected values, with these regions being tighter at an \ac{SNR} of eight.

What is wanted from a posterior which is largely the prior depends on how it will be used.
To follow up a faint signal with a telescope, the parts of the sky map which the data do not inform are unlikely to contain the source, and pointing at them is unlikely to be useful, so the prior-dominated part of the posterior is better ignored.
In a hierarchical analysis of a population the prior used during \ac{PE} for each event is divided out by reweighting to a population-informed prior, so that only the likelihood matters, but enough samples are then needed in the regions which the population prior favours~\citep{Essick:2022ojx,Talbot:2023pex}.
These two uses favour different treatments, and we return to this in section~\ref{sec:conclusion}.

Results are normally presented as one-dimensional marginal posteriors, obtained by integrating the joint posterior over the remaining parameters, and subsequently summarised by a median and a credible interval.
Summary statistics of this sort require a marginal distribution whose morphology makes them meaningful.
For example, the median of a distribution which has support over an open interval is meaningless.
In some prior-dominated posteriors, including those in figures~\ref{fig:gw200308} and~\ref{fig:gw2003222} they do not: the median and interval of a tail which follows the prior largely reflect the prior bounds.
For GW200308\_173609, \ac{GWTC3} quotes $D_L=7.1^{+13.9}_{-4.4}$\,Gpc, a $90\%$ interval from $2.7$ to $21$\,Gpc which is as much a statement about the prior as about the data~\citep{abbott_gwtc-3_2023}.

\subsection{Approach to quiet signals in the literature}

Catalogues of gravitational-wave signals, such as GWTC-5.0~\citep{catalog-gwtc-5}, which have appeared in the literature differ little in their default priors; they differ in what is done when the data do not dominate the posterior.
Two strategies are typically employed, sometimes in combination: restricting which events are analysed in order to remove problematic signals from consideration in the first place, and reporting the full posterior with caveats.

Many catalogues, including GWTC-1,-2,-4, and~-5, and the OGC catalogues~\citep{abbottGWTC1GravitationalWaveTransient2019,catalog-gwtc-2,catalog-gwtc-4d0,catalog-gwtc-5,catalog-1ogc,catalog-2ogc,catalog-3ogc,catalog-4ogc} perform full \ac{PE} only on candidates which pass a significance threshold, $p_\mathrm{astro}\geq0.5$ or a false alarm rate below once per hundred years.
\ac{GWTC4} and \ac{GWTC5} place a false alarm and $\pastro$ threshold~\citep{methods-gwtc-5d0}.
Sub-threshold candidates are released with the parameters of the template which triggered the search, but these represent a point estimate of the parameters of the signal, and do not provide posterior probability distributions.
In contrast GWTC-3 used a threshold of $p_\mathrm{astro}\geq0.5$ only, and as a result contains a larger number of marginal candidates.

This approach reduces the problem but does not remove it: candidates just above any threshold are still quiet, while information from candidates which do not exceed the threshold is not used.
~\citet{2026arXiv261116985W} examines the set of candidates which fail to meet the \ac{FAR} threshold in \ac{GWTC4}, but which exceed the $\pastro=0.5$ threshold.
They perform \ac{PE} on these candidates and identify the presence of prior-dominated posteriors in their results for two candidates.
We find that the posteriors~\citep{lurkers-datarelease} for three of these events have a significant percentage of samples with $\rhomf<\sqrt{8}$: GW230920\_064709 (1.9\%), GW230830\_064744 (1.6\%), and GW231102\_052214 (1.1\%).
Thus, even events which are more likely than not of astrophysical origin risk presenting this problem.

We summarise the criteria used in a selection of catalogue publications in table~\ref{tab:thresholds}.

\begin{table}
    \caption{Significance criteria for the candidates for which source parameters are reported in the catalogues discussed in the text.}
    \label{tab:thresholds}
    \begin{tabular}{lp{4.6cm}}
        \hline
        Catalogue & Criterion \\
        \hline
        4-OGC & $p_\mathrm{astro}>0.5$ or FAR $<1$ per $100$\,yr \\
        IAS (O3a) & $p_\mathrm{astro}>0.5$ \\
        \ac{GWTC3} & $p_\mathrm{astro}>0.5$ in any search, plus GW200105\_162426 \\
        \ac{GWTC4}, \ac{GWTC5} & FAR $<1$ per yr and $\pastro > 0.5$ in at least one search.  \\
        \hline
    \end{tabular}
\end{table}

For GW200308\_173609 and GW200322\_091133, which have inferred \acp{SNR} of $4.7$ and $4.5$ respectively, \ac{GWTC3} states that the high-mass, large-distance peak is dominated by the prior.
These candidates are indicated in the summary table and figures as being particularly prior sensitive, noting that the low-likelihood high-distance modes cannot be excluded and that a population-informed prior would alter the results.
However, the quoted credible intervals include the prior-dominated tail, for example $D_L = 7.1^{+13.9}_{-4.4}$\,Gpc is quoted for GW200308\_173609.
In the IAS analysis of O3a data a similar approach is taken for low-significance candidates.
GW190814\_192009 is noted for comparable peaks throughout a large region of parameter space~\citep{olsen_new_2022}.

Sampling with one prior chosen, and later reweighting the samples to another prior, has also been used in some catalogues, including GWTC$\leq3.0$, which draw samples from a posterior which had a prior which was uniform in Euclidean volume, and then reweighted to one which was uniform in co-moving volume~\citep{abac_gwtc-40_methods_2025}.
For a discussion of reweighting as a technique in \ac{GW} astronomy we direct the reader to~\citet{10.1093/rasti/rzag012}.
\citet{huang_characterization_2018} performed inference using that technique on simulated low-significance sources, using a prior uniform in distance and reweighting to uniform in volume, because a nested sampler with the volume prior would spend much of its time at large distance and, for a finite number of live points, might not explore the small-distance region properly.
They also find that the chirp-mass posteriors of simulated sources are frequently multimodal or very broad for an optimal network \ac{SNR} of seven or below, with a class of unimodal posteriors whose tails cover most of the prior range because the likelihood profile is shallow.
However, reweighting is only as good as the coverage of the baseline samples. 
\citet{yang_rapid_2026} sampled GW170817 with a prior $\propto\dl^2$ and reweighted to a prior uniform in $\dl$, which gives more weight to small distances. 
The reweighted posterior does not reproduce the one sampled with the uniform in $\dl$ prior directly: the $\dl^2$ prior gives little weight to a low-distance mode of the likelihood, so the baseline run has few samples there, and the reweighted posterior has a lower effective sample size than the baseline. 
In their application this shows as a missing high-$H_0$ tail, which the reweighting does not recover well. 
Sampling with a prior which favours small distances and reweighting to one which favours large distances, discussed in greater detail in section~\ref{sec:reweighting}, does not have this problem, but will still result in discarded samples.

A distance prior is also required for rapid localisation analyses. 
BAYESTAR~\citep{PhysRevD.93.024013} takes a power of distance, $p(r)\propto r^m$, with $m=2$ for a prior uniform in volume and $m=-1$ for one uniform in the logarithm of the distance. 
By default it uses the volume prior out to the horizon distance for a \ac{SNR} of four in the most sensitive detector, so that the upper bound is tied to the sensitivity of the detectors and not chosen by hand.

An alternative is to replace the default prior with one which is informed by the observed population~\citep{Moore:2021xhn,Galaudage:2019jdx}, or to reweight posteriors hierarchically~\citep{mandelExtractingDistributionParameters2019c,Thrane:2018qnx}. When done consistently these are the same approach: the population model is the prior for each event, and the two differ only in whether the events being reweighted are included in the population fit~\citep{2026PhRvD.114b1302M}. 
Population-informed analyses of this kind have been advocated for outlying \ac{GW} events~\citep{2020ApJ...891L..31F}, for catalogues in which the fitted population is used to inform new events~\citep{2026PhRvD.114b1302M}, and, outside \ac{CBC} analysis, for the noise parameters of individual pulsars in pulsar timing arrays, where uninformative priors on an ensemble bias the inferred gravitational-wave background~\citep{2024ApJS..273...23V}.
\ac{GWTC4} and \ac{GWTC5} provide plots of the marginal parameter distributions for their candidates both with the standard priors and reweighted to a population prior, however neither are used in the summary statistics presented in these works.

\subsection{Analogues outside \ac{CBC} analysis}\label{sec:analogues}

The difficulties described above are not unique to \acp{GW}, and other fields have met them.

Inferring a distance from a parallax $\varpi=1/D$ with Gaussian errors is effectively the same problem as ours, where the waveform amplitude is proportional to $1/\dl$.
\citet{bailer-jonesEstimatingDistancesParallaxes2015} shows that inverting the parallax becomes unreliable once the fractional error exceeds about $20\%$, equivalent here to a matched filter \ac{SNR} of $\rhomf\ la 5$.
They recommend a prior which decays smoothly at large distance, and note that for such a prior the posterior can have a data-dominated mode at small distance and a prior-dominated mode at large distance, as we see for GW200308\_173609 and GW200322\_091133.
They also warn that discarding objects above some fractional error wastes data and can bias subsequent analysis, and that bias corrections of the Lutz--Kelker type are valid only in restricted circumstances.
\footnote{This correction, from \citet{lutz_kelker_1973} concerns a sample of stars which are selected by a lower limit on their observed parallax.
The volume grows with distance, so there are more stars with a small parallax than large, so the observed parallaxes are, on average, larger than the true ones, and they derived corrections for the absolute magnitudes inferred from them.
For this parallax model the information for the distance is $\propto r^{-4}$, so the Jeffreys prior is $\propto r^{-2}$~\citep{ghosh_curious_2024}, which is uniform in the parallax $1/r$.}

Astronomical surveys face selection and measurement error together.
\citet{loredo_multilevel_2019}, building on~\citet{loredoAccountingSourceUncertainties2004}, shows that when the data used to characterise a source are also used to detect it, the detection efficiency enters the population likelihood through the expected number of detections and not as a factor on each source's likelihood or prior: inserting it there would use some of the data twice, corrupting the inference.
In ground-based \ac{CBC} analysis the same strain data are used for both detection and \ac{PE}, so the first case applies.
They also advocate reporting the likelihood function of each source rather than a best-fit value with an uncertainty, so that marginal detections and limits can be used.
For fixed values of the other parameters, the likelihood in $1/\dl$ is described by only two numbers, $\rhomf/\rho_1$ and $1/\rho_1$.

Weakly constrained parameters are routinely encountered in particle physics.
\citet{demortier_reference_2010} notes that uniform priors depend on the parametrisation and can give improper posteriors, for example for upper limits on cross sections, and advocate reference priors, which are invariant under reparametrisation and give proper posteriors.
They caution that for small samples the results should be treated as preliminary, and the sensitivity to the prior examined.
\citet{feldman_unified_1998} show that choosing between an upper limit and a two-sided interval after seeing the data breaks the stated coverage of the intervals, and propose a unified construction; they also recommend reporting the experiment's sensitivity alongside a limit.
Profile likelihoods are also widely used within particle physics~\citep{cowanAsymptoticFormulaeLikelihoodbased2011}.
These are an alternative to the marginal posterior which uses only the likelihood: the nuisance parameters $\boldsymbol{\theta}_{-i}$ are maximised over rather than marginalised over:
\begin{equation}
    \mathcal{L}_\mathrm{prof}(\theta_i) = \max_{\boldsymbol{\theta}_{-i}} \mathcal{L}(\theta_i, \boldsymbol{\theta}_{-i}).
\end{equation}
As these are constructed purely from the likelihood function they do not depend on the prior.

In targeted searches for continuous waves, \citet{prix_targeted_2009} demonstrates that the F-statistic is the Bayes factor for a prior distribution which is uniform in the amplitude parameters.
In terms of the physical amplitude $h_0$ this is proportional to $h_0^3$, and so strongly favours loud signals.
Instead they propose the B-statistic, which uses a prior which is uniform in $h_0$.
Therefore using a frequentist method does not automatically circumvent the potential issues which arise from prior choices, as these prior choices may still be implicitly encoded within the statistic.

\section{Standard distance priors in \ac{GW} analysis}\label{sec:distance-priors}

The most straight-forward choice of prior over the distance is to assume that detected events should be distributed uniformly through Euclidean space, and that the Universe is both flat and is not expanding.
These approximations are reasonable at low redshifts $z\ll 1$, and can be encoded in a power-law prior with $\alpha=2$:
\begin{equation}\label{eq:powerlaw}
p(\dl) = \frac{(1+\alpha) \dl^\alpha}{B^{1+\alpha} - A^{1+\alpha}} 
\end{equation}
where $\dl \in [A, B]$ represent the lower and upper limits of the prior.

This distance prior was widely used in early \ac{GW} analysis, where detectors were only sensitive to events at low redshift, and was the prior used for sampling in catalogue analyses up to and including \ac{GWTC3}~\citep{abac_gwtc-40_methods_2025}. 
Since GWTC-2 the published posteriors have also been reweighted to a prior uniform in co-moving volume and source-frame time~\citep{abbott_gwtc-3_2023}.
As detectors have become more sensitive it has become possible to detect signals from distances where the assumptions of non-expansion break-down.
Since hierarchical analyses of sources require evaluating the ratio of the astrophysical population prior to the default prior, a large discrepancy between these reduces the number of effective samples from any particular event, increasing the statistical error in population analysis~\citep{Essick:2022ojx, Talbot:2023pex}.

In \ac{GWTC4} the analyses of new candidates therefore sample directly from the prior which is uniform in co-moving volume and co-moving time, to which the earlier results were reweighted:
\begin{equation}\label{eq:uniform-frame}
p(z) \propto \frac{1}{1 + z}\frac{dV_{c}}{dz}
\end{equation}
This prior assumes that sources are uniformly distributed in co-moving (four)-volume, accounting for the time dilation between the observer's frame and the source frame through the $(1+z(\dl))^{-1}$ factor.
Sampling with this prior in \ac{GWTC4}~\citep{catalog-gwtc-4d0,abac_gwtc-40_methods_2025} allows for the inclusion of events at higher redshift in the detectors' more sensitive state during their fourth observing run.

We could alternatively consider a prior which is uniform only in the co-moving (space) volume:
\begin{equation}\label{eq:uniform-comoving}
p(z) \propto \frac{d V_{c}}{dz}
\end{equation}
though it is difficult to justify this choice over the choice of the uniform-in-source-frame prior of equation~\ref{eq:uniform-frame}.

The priors above differ in the redshift evolution of the merger rate which they implicitly assume.
For the prior in equation~\ref{eq:uniform-frame} the corresponding rate density of mergers is constant in the source frame.
For equation~\ref{eq:uniform-comoving} the merger rate density is constant in co-moving volume and per-unit of observer time, and therefore increases as $(1+z)$ in the source frame.
Neither incorporates the evolution expected if the rate follows the lagged star formation rate.

\subsection{Effect of the prior on the matched-filter threshold}\label{sec:cosmo-threshold}

In section~\ref{sec:problem} we identified that for the Euclidean prior there is a threshold $\rhomf=\sqrt{8}$ below which the posterior has no local maximum in $\dl$ and is dominated by the prior at large distance.
For a prior $\propto\dl^\alpha$ the stationary points of the posterior in $\dl$, for fixed values of the other parameters, satisfy $\alpha\dl^2-\rhomf\rho_1\dl+\rho_1^2=0$, so that the posterior has a finite mode, $\dl^\mathrm{MAP}=\rho_1\left[\rhomf-\sqrt{\rhomf^2-4\alpha}\right]/(2\alpha)$, only if $\rhomf\geq2\sqrt{\alpha}$, which is $\sqrt{8}$ for $\alpha=2$.

Priors such as the uniform-in-comoving-volume prior are not power laws, but the same condition holds at the mode with $\alpha$ replaced by the local logarithmic slope of the prior,
\begin{equation}
    \alpha_\mathrm{eff}(\dl)=\mathrm{d}\ln p/\mathrm{d}\ln\dl,
\end{equation}
and where this varies slowly the threshold is $2\sqrt{\alpha_\mathrm{eff}}$
At the threshold the mode is at $\dl=\rho_1/\sqrt{\alpha_\mathrm{eff}}$, where the optimal \ac{SNR} of the source is $\sqrt{\alpha_\mathrm{eff}}$.
In table~\ref{tab:cosmo-threshold}, we show the values of $\alpha_\mathrm{eff}$ and the corresponding threshold for the two cosmological priors of this section.

\begin{table}
\caption{
    The local logarithmic slope $\alpha_\mathrm{eff}$ of the prior on $\dl$, and the threshold $2\sqrt{\alpha_\mathrm{eff}}$ on $\rhomf$ for a mode at that distance, for the prior uniform in comoving volume and source-frame time (equation~\ref{eq:uniform-frame}) and for the prior uniform in comoving volume (equation~\ref{eq:uniform-comoving}), with the Planck15 cosmology. The Euclidean prior has $\alpha=2$ and a threshold of $\sqrt{8}\approx2.83$.}
\label{tab:cosmo-threshold}
\centering
\begin{tabular}{rrrrrr}
\hline
 & & \multicolumn{2}{c}{source frame} & \multicolumn{2}{c}{comoving} \\
$\dl$ [Mpc] & $z$ & $\alpha_\mathrm{eff}$ & $2\sqrt{\alpha_\mathrm{eff}}$ & $\alpha_\mathrm{eff}$ & $2\sqrt{\alpha_\mathrm{eff}}$ \\
\hline
$100$ & $0.02$ & $1.89$ & $2.75$ & $1.91$ & $2.77$ \\
$300$ & $0.07$ & $1.71$ & $2.62$ & $1.77$ & $2.66$ \\
$1000$ & $0.20$ & $1.29$ & $2.27$ & $1.44$ & $2.40$ \\
$2000$ & $0.36$ & $0.92$ & $1.92$ & $1.15$ & $2.15$ \\
$4000$ & $0.65$ & $0.49$ & $1.39$ & $0.81$ & $1.80$ \\
$8000$ & $1.1$ & $-0.01$ & --- & $0.42$ & $1.29$ \\
\hline
\end{tabular}
\end{table}

The threshold is therefore lower than $\sqrt{8}$ and decreases with distance.
This helps to demonstrate why analysis of the two low-significance candidates GW200308\_173609 and GW200322\_091133, which have $\rhomf=4.7$ and $4.5$, respectively, with the prior uniform in source-frame volume and time gives a posterior with a mode at finite distance, while the same analysis with the Euclidean prior does not.
The long tail is not visible in the posterior samples reweighted to the uniform-in-source-frame prior.

\section{The Malmquist prior}\label{sec:malmquist}

A natural approach is to appeal to the notion of a \emph{Malmquist bias}, accounting for the difficulty in detecting signals which are weak relative to those which are strong.
A signal may be weak either because it is intrinsically faint, or simply because it is far away, since the amplitude of the strain is inversely proportional to the distance to the source.
Any experiment which is flux-limited in some way will therefore be biased towards the detection of intrinsically loud sources, which can be seen to larger distances.
Since this bias acts on the distance and the intrinsic loudness of the source jointly, it is tempting to build it into the prior on $\dl$.

\begin{equation}
p_M(\dl, \theta|\mathrm{det}) \propto P(\mathrm{det}|\dl, \theta)p(\dl, \theta)
\label{eq:malmquist}
\end{equation}

Indeed, a comparable approach has been implemented in LALInference~\citep{code-lalinference} which multiplies the prior by an indicator function of the optimal \acp{SNR} of the template waveform.
A sample is rejected unless its \ac{SNR} in the second-loudest detector exceeds a threshold, and the loudest-detector and network \acp{SNR} are each above some specified threshold.

This represents the case of equation~\ref{eq:malmquist} where $P(\mathrm{det}|\dl,\theta)$ is a step function at the horizon distance $D_h(\theta)$.
The requirement for \ac{SNR} in the second-loudest detector represents a coincidence condition rather than a threshold on the network \ac{SNR}.

If a candidate is determined to be detected when its \ac{SNR} exceeds some threshold $\rho_\mathrm{th}$, then the probability of detection for a source with parameters $\theta$ and distance $\dl$ depends on the statistic used by the search.

Searches maximise over the phase of the signal, so that in each of $N_\mathrm{det}$ detectors the squared \ac{SNR} is the sum of the squares of two independent unit-variance Gaussian variables, and the network $\rho^2$ follows a non-central $\chi^2$ distribution with $2N_\mathrm{det}$ degrees of freedom and non-centrality $\rho_1(\theta)^2/\dl^2$.

The probability of detection is then
\begin{equation}\label{eq:pdet}
    P(\mathrm{det} | \dl, \theta) = Q_{N_\mathrm{det}}\!\left(\frac{\rho_1(\theta)}{\dl},\,\rho_\mathrm{th}\right),
\end{equation}
where $Q_M$ is the generalised Marcum $Q$-function, and $N_\mathrm{det}=2$ for the two detectors used here~\citep{2024CQGra..41l5002G}. At large optimal \acp{SNR} the posterior approaches the Gaussian approximation, and the selection function is approximately $P\approx\frac{1}{2}\mathrm{erfc}[(\rho_\mathrm{th}-\rho_1(\theta)/\dl)/\sqrt{2}]$, in which the matched-filter \ac{SNR} is normally distributed with unit variance about the optimal \ac{SNR}.
This is a smooth step function, falling from one to zero around the horizon distance $D_h(\theta) = \rho_1(\theta)/\rho_\mathrm{th}$.

\begin{figure}
    \centering
    \includegraphics[width=\linewidth]{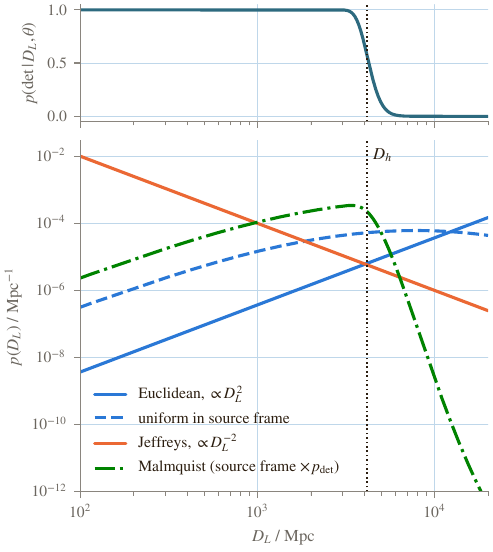}
    \caption{
        Priors on the luminosity distance for a single non-spinning source with detector-frame chirp mass $30\,M_\odot$, evaluated with the Advanced LIGO design PSD~\citep{ligo-t1800044}. 
        Top: the detection probability from equation~\ref{eq:pdet} for $\rho_\mathrm{th}=8$, with the horizon distance $D_h$ marked (dotted line). Bottom: the normalised priors on $D_L \in [100, 20000]\,\mathrm{Mpc}$: Euclidean ($\propto D_L^2$), uniform in source-frame comoving volume and time, Jeffreys ($\propto D_L^{-2}$ for this source, shown normalised), and the Malmquist prior from equation~\ref{eq:malmquist} built from the source-frame prior.
        }
    \label{fig:prior-comparison}
\end{figure}

When used in equation~\ref{eq:malmquist} this has the effect of removing the long tail at large distance, since the prior is exponentially suppressed beyond $D_h(\theta)$.
Because $D_h$ depends on the chirp mass, inclination and sky location of the source, the prior on $\dl$ is necessarily coupled to the prior on these parameters, and the joint prior cannot be written as a product of marginals, $p(\theta,\dl)\not\propto p(\theta)p(\dl)$. This Malmquist prior is shown in figure~\ref{fig:prior-comparison}, for a fixed non-spinning source with a chirp mass of $30\,M_\odot$ and a threshold of $\rho_\mathrm{th}=8$.
For this source the Euclidean and source-frame priors place 99\% and 88\% of their mass beyond the horizon distance, where the source could not have been detected, while the Malmquist prior places 13\% there.
We assume a two-detector Hanford--Livingston network with the Advanced \ac{LIGO} design \ac{PSD}~\citep{ligo-t1800044} and a lower frequency cutoff of $20$\,Hz; the sky location ($\alpha=5^\mathrm{h}15^\mathrm{m}8^\mathrm{s}$, $\delta=-69.4^\circ$) and inclination ($\theta_{JN}=0.4$\,rad) are fixed, which also fixes $\rho_1(\theta)$ and thus $D_h$.

\subsection{Conditioning on detection}

A detection is a statement about the data: the candidate was identified because $d$ produced a statistic exceeding some threshold: $\mathrm{det}(d) = 1$.
Conditioning both the likelihood and the prior on detection gives a likelihood with the form
\begin{equation}
    p(d | \theta, \mathrm{det}) = \frac{p(d | \theta)\,\mathbf{1}[\mathrm{det}(d)]}{p(\mathrm{det} | \theta)},
\end{equation}
and a prior
\begin{equation}
    p(\theta | \mathrm{det}) \propto p(\mathrm{det} | \theta)\, p(\theta).
\end{equation}

The factors of $p(\mathrm{det}|\theta)$ cancel in the product, so the posterior has the form
\begin{equation}
    p(\theta | d, \mathrm{det}) \propto p(d | \theta)\, p(\theta),
\end{equation}
which is the ordinary posterior, since the indicator function has a value of one for the data which we have in fact observed and therefore analysed.
A consistent treatment of the selection therefore leaves the single-event posterior unchanged.
For a more detailed discussion of conditioning on the source parameters rather than data see~\citet{2024ApJ...962..169E}.
The selection function does matter in population inference, where it enters through the normalisation of the population model~\citep{loredoAccountingSourceUncertainties2004,mandelExtractingDistributionParameters2019c,vitaleInferringPropertiesPopulation2021}.
There it constrains the hyperparameters of the population model, and hence the population distribution which serves as the prior for each event.
However, it does not appear as an additional factor in the single-event prior or likelihood: the prior is the intrinsic population, and not the population of detected sources.

The Malmquist prior from equation~\ref{eq:malmquist} differs from this: it retains the factor $p(\mathrm{det}|\theta)$ in the prior while using the unmodified likelihood $p(d|\theta)$.
The posterior is then proportional to $p(d|\theta)\,p(\mathrm{det}|\theta)\,p(\theta)$, in which the detection is accounted for twice: once through the data $d$ which the likelihood evaluates, and again through the weighting of the prior.
In this sense the Malmquist prior double-counts the data.
This problem is also found in astronomical surveys~\citep{loredo_multilevel_2019}, and it therefore should not be thought of as a principled correction of the posterior, but as a regularising choice which happens to suppress the high-distance tail.
Its practical disadvantages are that it depends on the (pipeline-specific) detection threshold and on the noise \ac{PSD} at the time of the event, and that it is $\theta$-dependent as described above.

\section{Jeffreys priors}\label{sec:jeffreys}

The general approach taken in \ac{CBC} parameter estimation asserts that the prior should be informative, and ought to reflect our prior understanding of the Universe.
As such we should expect to construct our prior on a quantity such as the luminosity distance of a signal taking into account our knowledge of the geometry of the Universe.
However, a different approach is also available to us, which simply considers the information we can gain from the likelihood.
Such a prior is known as a \emph{reference prior} (see discussion in~\citep{berger_formal_2009}; we note that some \ac{GW} literature uses this term for the default \ac{PE} prior~\citep{2026PhRvD.114b1302M}), and in the asymptotic limit of a large number of independent observations (for a \ac{GW} signal, the high-\ac{SNR} limit) this is the Jeffreys prior (see~\citep{jeffreys1998theory}), as shown in~\citep{clarkeJeffreysPriorAsymptotically1994}.

We construct such a prior using the information matrix, $\Gamma_{ij}$, which represents the information available about the parameters $\theta$ in the linearised signal approximation~\citep{Cutler:1994ys,Vallisneri:2007ev}, and is defined 
\begin{equation}
    \label{eq:information-matrix}
    \Gamma_{ij}(\vec{\theta}) = \left\langle \frac{\partial h}{\partial \theta^i} \middle| \frac{\partial h}{\partial \theta^j} \right\rangle.
\end{equation}

Considering the luminosity distance, $\dl$, such that the waveform amplitude $h \propto \dl^{-1}$ we have 
\begin{equation}
    \Gamma_{\dl \dl} = \left\langle \frac{\partial h}{\partial \dl} \middle| \frac{\partial h}{\partial \dl} \right\rangle = \frac{1}{\dl^2} \langle h | h \rangle = \frac{\rho^2}{\dl^2}.
\end{equation}
Where all other parameters are fixed, we have $\rho \propto \dl^{-1}$ and so $\Gamma_{\dl \dl} \propto \dl^{-4}$. 

If we determine the optimal \ac{SNR} at $1$\,Mpc as a reference, which we call $\rho_1(\theta)$ for any given $\theta$, then $\rho=\rho_1(\theta)/\dl$.
We can then express the appropriate component of the information matrix as
\begin{equation}
    \Gamma_{\dl\dl}=\rho_1^2(\theta)/\dl^4.
\end{equation}

The Jeffreys prior is defined as 
\begin{equation}
    p(\theta) \propto \sqrt{\det \Gamma_{ij}(\theta)}.
\end{equation}
Were we to use the Jeffreys prior, joint over all parameters of the problem we would likely find it too uninformative to be useful.
However, the prior does scale as $D_L^{-1}$ in all its entries, and $D_L^{-2}$ in those for $D_L$ specifically.
So if we condition on the other parameters we have, for $\dl\in[d_{\rm min},d_{\rm max}]$:
\begin{equation}
    \label{eq:jeffrey-dl}
    p_J(\dl\,|\,\theta) = \frac{\rho_1(\theta)}{\dl^2} = \frac{\rho}{\dl}.
\end{equation}

\begin{figure*}
    \centering
    \includegraphics[width=\textwidth]{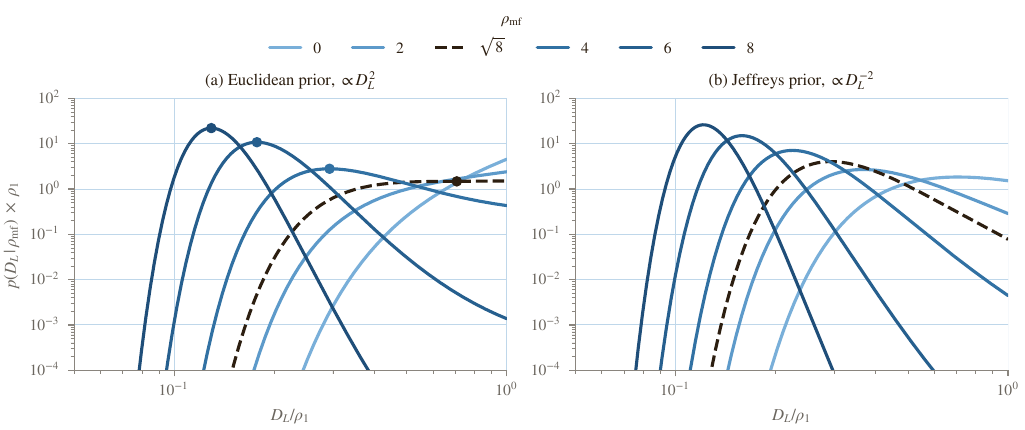}
    \caption{
        The $\rhomf=\sqrt{8}$ threshold.
        (a) The posterior on $\dl/\rho_1$ for fixed values of the other parameters, a Gaussian likelihood in $1/\dl$ and a prior $\propto\dl^2$ over $0.05\leq\dl/\rho_1\leq1$, for several values of $\rhomf$.
        The dots mark the local mode where one exists, which requires $\rhomf\geq\sqrt{8}$; at $\rhomf=\sqrt{8}$ (dashed) and below, the posterior has no local maximum and increases to the upper bound. (b) The same likelihood with the Jeffreys prior of equation~\ref{eq:jeffrey-dl}, which is $\propto\dl^{-2}$ at fixed values of the other parameters; the posterior has a local maximum for every $\rhomf$.}
    \label{fig:sqrt8}
\end{figure*}

This prior is flat in the optimal \ac{SNR} $a=\rho_1(\theta)/\dl$, with $p(a \mid \theta)=\mathrm{const.}$ for any given $\theta$. 
We refer to this as the conditional Jeffreys prior over distance.
Unlike the Euclidean prior, this prior does not produce a threshold in $\rhomf$.
At fixed $\theta$ the posterior density in $\dl$ has a local maximum at $\dl/\rho_1=\left(\sqrt{\rhomf^2+8}-\rhomf\right)/4$, which exists for every value of $\rhomf$, including $\rhomf=0$, compared with $\left(\rhomf-\sqrt{\rhomf^2-8}\right)/4$ for the Euclidean prior, which exists only for $\rhomf\geq\sqrt{8}$ (see figure ~\ref{fig:sqrt8}).

This form of the prior follows the construction of a reference prior, in which a joint prior is built in stages~\citep{berger_formal_2009}.
The first stage is the prior on $\dl$ conditional on the other parameters $\theta$, chosen to be the one that lets the data tell us the most about $\dl$ when $\theta$ is held fixed.
The remaining stage, the priors on the other components of $\theta$, are left to us, and we take it from our astrophysical knowledge, as for any other analysis.
The reference prior is then used only for $\dl$ given $\theta$, and the joint prior is the product of this and the other priors.

At fixed $\theta$ the data depend on $\dl$ only through the matched-filter \ac{SNR}, and from the likelihood of section~\ref{sec:marginal-jeffrey} this is distributed as $\rhomf\sim\mathcal{N}(a,1)$, where $a=\rho_1(\theta)/\dl$ is the optimal \ac{SNR}.
So the data measure $a$ with unit uncertainty, and $a$ is confined by the limits on distance to a range of width $\Lambda=\rho_1(\theta)(d_{\min}^{-1}-d_{\max}^{-1})$.
The information gained about $a$ can be quantified by the Kullback--Leibler divergence between the posterior and the prior, averaged over the data we might observe~\citep{bernardo_reference_1979}, and the reference prior is the prior for which this is greatest.
If $\Lambda\gg1$, so that the range of $a$ is wide compared with the unit uncertainty of the measurement, this is the prior flat in $a$, which is the prior of equation~\ref{eq:jeffrey-dl}, as in the general result of~\citet{clarkeJeffreysPriorAsymptotically1994}.
Intuitively, every value of $a$ is then equally well measured, so none should be favoured.
For smaller $\Lambda$ it is not exactly optimal, because the best prior is then a set of discrete points~\citep{smith_information_1971}, but the information lost by using the flat prior is small: $0.21$, $0.16$, $0.10$ and $0.06$\,nat for $\Lambda=4$, $8$, $16$ and $32$, while the information gained is itself about $\ln\Lambda-1.4$\,nat for large $\Lambda$.
This condition is on the width of the prior range and not on the \ac{SNR} of the signal we observe.

The prior is optimal for the information on $\dl$ if $\theta$ were known, whatever the prior on $\theta$.
It is not optimal for the information on $\dl$ when $\theta$ is unknown, which is the case in an analysis, nor for the joint information on $(\theta,\dl)$, and parameters which are degenerate with the distance, such as the inclination, are where we would expect the difference to matter most.
The statement also applies to the conditional prior normalised over $[d_{\min},d_{\max}]$, which is flat in $a$ and does not depend on $\rho_1$.
The factor $\rho_1(\theta)$ in equation~\ref{eq:jeffrey-dl} is not fixed by it, and instead weights the prior on $\theta$ (section~\ref{sec:marginal-jeffrey}).

This form of the prior has a radically different form from the conventionally-used prior, as can be seen visually when compared to the uniform in comoving distance and Malmquist priors in figure~\ref{fig:prior-comparison}, and it will strongly favour close-by signals, and will disfavour distant, weak signals.

\begin{figure*}
    \centering
    \includegraphics[width=\textwidth]{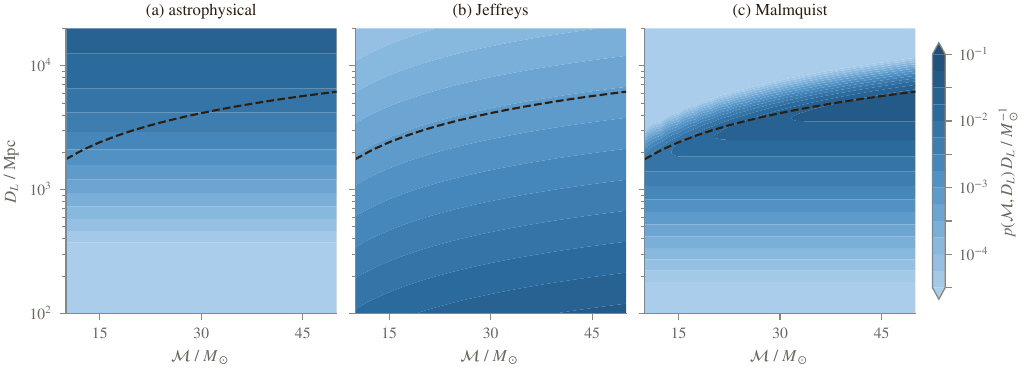}
    \caption{
    Joint priors on the chirp mass and luminosity distance, shown as the density per unit $\Mc$ and per unit $\ln\dl$ and normalised over the whole plane in each panel, for a chirp-mass prior $\pi(\Mc)$ uniform over $10$--$50\,M_\odot$ and the other parameters as in figure~\ref{fig:prior-comparison}. (a) The uniform in comoving volume prior (b) The conditional Jeffreys prior from equation~\ref{eq:jeffrey-dl}, which gives more prior weight at a given distance to loud, high-mass sources and the same weight to every source at a given \ac{SNR}. (c) The Malmquist prior from equation ~\ref{eq:malmquist} with $\rho_\mathrm{th}=8$, which both removes the large distances and favours high masses. The dashed line is the horizon distance $D_h(\Mc)$.}
    \label{fig:prior-joint}
\end{figure*}

The prior of equation~\ref{eq:jeffrey-dl} depends on the other parameters through the factor $\rho_1(\theta)$, which sets the range of optimal \ac{SNR}, $\rho_1/d_{\max}$ to $\rho_1/d_{\min}$, that the fixed range of distance covers.
In figure~\ref{fig:prior-joint} we show the priors in the plane of chirp mass and distance. 
Where the uniform-in-source-frame prior has no dependence on the chirp mass, both the Jeffreys and Malmquist priors do, via the dependence of \ac{SNR} on the chirp mass.

\subsection{Marginalisation of distance Jeffreys prior}
\label{sec:marginal-jeffrey}

The prior presented in equation~\ref{eq:jeffrey-dl} has an additional attractive feature: we can analytically marginalise over $\dl$.
Considering the definition of the likelihood in equation~\ref{eq:loglikelihood}, and changing variables to $u = \dl^{-1}$, the conditional Jeffreys prior transforms as $p(u) = p(\dl)\,|\mathrm{d}\dl/\mathrm{d}u| \propto u^{2}\cdot u^{-2}$: the Jacobian cancels the prior exactly, so $p(u)$ is uniform on $[d_{\max}^{-1}, d_{\min}^{-1}]$, with density $\rho_1(\theta)$. The likelihood is
\begin{align}
    \mathcal{L} &\propto \exp\left(\rhomf\rho_1 u - \frac{1}{2}\rho_1^2u^2\right) \nonumber \\
    &= e^{\rhomf^2/2}\exp\left[-\frac{1}{2}\rho_1^2\left(u - \rhomf/\rho_1\right)^2\right],
\end{align}
which is Gaussian in $u$ with mean $\rhomf/\rho_1$ (the inverse of the maximum-likelihood distance) and standard deviation $1/\rho_1$. The posterior $p(u|d)$ is therefore a truncated normal distribution, and the marginal likelihood is thus
\begin{align}
    \mathcal{L}_{\rm marg} = & e^{\rhomf^2/2}\sqrt{\frac{\pi}{2}} \nonumber \\
    & \cdot \left[ \mathrm{erf}\!\left(\frac{\rho_1/d_{\min}-\rhomf}{\sqrt2}\right) 
        - \mathrm{erf}\!\left(\frac{\rho_1/d_{\max}-\rhomf}{\sqrt2}\right)\right],
    \label{eq:LJ}
\end{align}
up to the constant $e^{c}$. Given $\rho_1$ and $\rhomf$ for a set of the remaining parameters, posterior samples of distance are reconstructed by drawing $u$ from the truncated normal and setting $\dl = 1/u$.

The same change of variable removes the large-distance divergence of the integrand in the detection statistic used by~\citet{olsen_new_2022}.
There the prior uniform in volume gives a factor $a^{-4}$ in terms of the optimal \ac{SNR} $a=\rho_1/\dl$, whereas $\dl^{-2}\,\mathrm{d}\dl\propto\mathrm{d}a$, so that under the conditional Jeffreys prior the integrand is finite as $a\rightarrow0$ and the integral does not depend on an upper cutoff in distance chosen by hand.

\section{Reweighting to astrophysical priors}\label{sec:reweighting}

The nested samples from the analysis with the conditional Jeffreys prior can be reweighted to the astrophysical prior without a further sampling stage.
Let $\theta$ denote all of the parameters other than $\dl$, and let $L(\theta,\dl)$ be the likelihood.
The analysis using the conditional Jeffreys prior provides samples $\theta_i$ with weights $w_i\propto L_J(\theta_i)\,\Delta X_i$, where $\Delta X_i$ is the prior volume associated with the $i$th sample, measured under the prior $\pi(\theta)$ on the other parameters, and $L_J$ is the marginal likelihood shown in equation~\ref{eq:LJ}, that is
\begin{equation}
    L_J(\theta) = \int L(\theta,\dl)\,p_J(\dl\,|\,\theta)\,\mathrm{d}\dl ,
\end{equation}
where $p_J$ is the Jeffreys prior of equation~\ref{eq:jeffrey-dl}, which also depends on $\theta$ through $\rho_1(\theta)$.
The posterior for the astrophysical prior $p_T(\dl)$ has the marginal distribution
\begin{equation}
    p_T(\theta|d)\propto L_T(\theta)\,\pi(\theta),
\end{equation}
with
\begin{equation}
    \qquad L_T(\theta)=\int L(\theta,\dl)\,p_T(\dl)\,\mathrm{d}\dl,
\end{equation}
so the samples $\theta_i$ with weights $w_i' \propto w_i\,L_T(\theta_i)/L_J(\theta_i)$ represent it.
Given $\theta_i$, the distance has the conditional posterior $p_T(\dl|\theta_i,d)\propto L(\theta_i,\dl)\,p_T(\dl)$, which can be sampled directly.
The pairs $(\theta_i,\dl)$ are therefore weighted samples from the joint posterior for the astrophysical prior, with no further sampling or sequential Monte Carlo stage, provided that
\begin{itemize}
    \item the Jeffreys analysis posterior has support wherever the astrophysical posterior does, including over the same range of $\dl$;
    \item  the prior on the other parameters is the same in both analyses and does not depend on $\dl$, as is the case for priors on the detector-frame masses (a prior on the source-frame masses couples to $\dl$ through the redshift, and would have to be included in the integral defining $L_T$); and
    \item the likelihood scales exactly as $\dl^{-1}$, which excludes phase marginalisation in the Jeffreys analysis.
\end{itemize}

The weights are the ratio of two integrals of the same likelihood, and not the ratio of the priors $p_T/p_J\propto\dl^{4}$ evaluated at sampled distances, which grows by a factor of $\dl^4$ across the prior range.

\subsection{Efficiency of the reweighting}\label{sec:reweight-efficiency}

\begin{figure}
    \centering
    \includegraphics[width=\linewidth]{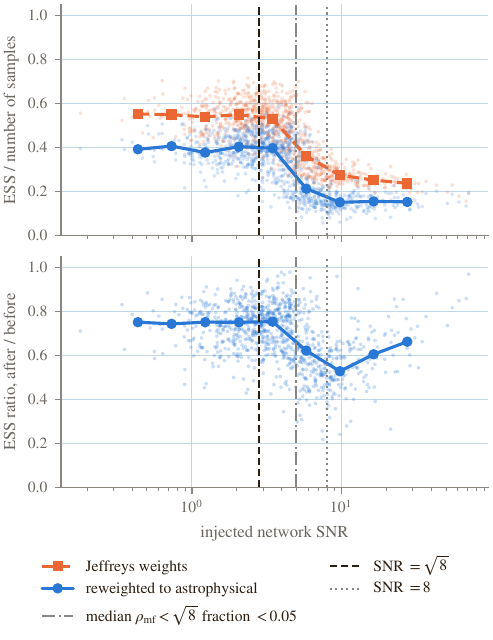}
    \caption{The effective sample size (ESS) of the nested samples against the injected network \ac{SNR}, for the $1024$ injected signals. Top: the ESS as a fraction of the number of samples, with the Jeffreys weights for equation~\ref{eq:jeffrey-dl} (squares) and after reweighting to the astrophysical prior (circles), with the median in logarithmic \ac{SNR} bins. Bottom: the ratio of the Jeffreys and astrophysical ESS from the top plot, showing the efficiency of the reweighting process. The dotted line marks a network \ac{SNR} of eight, the dashed line a network \ac{SNR} of $\sqrt{8}$, and the dash-dotted line a network \ac{SNR} of five, above which the median fraction of the Jeffreys posterior at $\rhomf<\sqrt{8}$ is below $0.05$}.
    \label{fig:efficiency}
\end{figure}

We measured the efficiency of this step with the \ac{ESS} of the weights, which we show in figure~\ref{fig:efficiency} for $1024$ injections drawn from the uniform in co-moving volume prior, but then analysed using the conditional Jeffreys prior on the luminosity distance, and the marginalisation scheme described in section~\ref{sec:marginal-jeffrey}.
For signals with a network \ac{SNR} below four the median \ac{ESS} was $54\%$ of the samples before reweighting and $40\%$ after, so that a median of $75\%$ of the effective samples are retained.
At higher \ac{SNR} both of these fall, as the posterior becomes narrow compared with the prior, and above an \ac{SNR} of $16$ to median \ac{ESS} is $24\%$ before and $15\%$ after reweighting above an \ac{SNR} of sixteen.
The lowest fraction of samples retained, $55\%$, was for the subset of signals with \acp{SNR} between six and $16$.
Above \ac{SNR} of $16$ a median of $66\%$ of samples were retained.

\begin{figure}
    \centering
    \includegraphics[width=\linewidth]{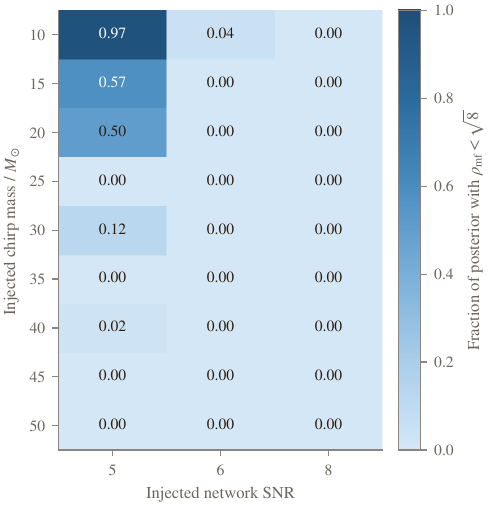}
    \caption{The fraction of posterior samples with a network matched-filter \ac{SNR} below $\sqrt{8}$, for each of the simulated injections.}
    \label{fig:sea-fraction}
\end{figure}

The reduction in the \ac{ESS} when using the conditional Jeffreys prior corresponds to the loss of the region of the posterior where $\rhomf<\sqrt{8}$, as can be seen in figures~\ref{fig:sqrt8-events} and~\ref{fig:sea-fraction}.
Around $10\%$ of the posterior samples lie in this region for injections with network \acp{SNR} between four and five, and $1\%$ between five and six, and we see a commensurate drop in the median \ac{ESS}.

\subsection{Validation against direct sampling}

We checked the method against direct sampling for four of the injections, with network \acp{SNR} of about 1, 3, 5 and 7.
For each we ran a second analysis which sampled $\dl$ explicitly under the astrophysical prior where the other parameters are independent of $\dl$, with the same data and with the same sampler settings as the analyses with the marginalised likelihood.
We compared these to the analyses sampled with the likelihood normalised over the distance, and then reweighted to the comoving volume prior.
Over all $15$ parameters the Jensen--Shannon divergence between the reweighted and the direct marginal posteriors is at most $0.05$\,nats (for the right ascension of the \ac{SNR}-seven signal), estimated from histograms of $2000$ resampled points in $30$ bins.

These analyses sampled $\dl$ explicitly, whereas the a more standard approach uses bilby's numerical marginalisation over distance~\citep{PhysRevD.93.024013}.
Performing an analysis with this marginalisation, we find the Jensen--Shannon divergence between the reweighted and the bilby marginal posteriors is at most $0.04$\,nats (again the right ascension of the \ac{SNR}-seven signal).

\section{Testing the conditional Jeffreys prior in \ac{PE}}\label{sec:verification}

\subsection{Injections}\label{sec:injections}

In order to investigate the situations in which parameter estimation will produce results which exhibit prior dominance we performed a study in which simulated signals were injected into simulated detector noise.

We simulated Gaussian noise with the Advanced LIGO design \ac{PSD}~\citep{ligo-t1800044} in two detectors, and injected non-spinning signals with the \texttt{IMRPhenomXPHM}~\citep{PhysRevD.103.104056} model at a range of chirp masses and network \ac{SNR} values.
The chirp mass was varied from $10$ to $50\,M_\odot$ in steps of $5\,M_\odot$ and the network \ac{SNR} was set to five, six or eight by scaling the luminosity distance.
The remaining parameters were the same for every signal, and are given in table~\ref{tab:fixed-params}.
The chirp mass is the detector-frame value, with no redshift applied to the injected masses, and the chirp-mass prior is on the detector-frame chirp mass.
The signals were analysed with \texttt{bilby}~\citep{code-bilby}, with a luminosity-distance prior uniform in comoving volume and source-frame time (Planck 2015 cosmology) over $10$--$30\,000$\,Mpc, a chirp-mass prior uniform in the component masses over $0.5$--$1.5$ times the injected value, and the mass ratio bounded to $0.05<m_2/m_1\leq 1$.

\begin{table}
    \caption{Parameters held fixed across the injections. The chirp mass ($10$--$50\,M_\odot$) and the network \ac{SNR} (5, 6, 8) were varied.}
    \label{tab:fixed-params}
    \begin{tabular}{ll}
        \hline
        Parameter & Value \\
        \hline
        Mass ratio $q$ & $1$ \\
        Component spins & $0$ \\
        Inclination $\iota$ & $0.4$\,rad \\
        Polarisation angle $\psi$ & $2.659$\,rad \\
        Phase $\phi$ & $1.3$\,rad \\
        Right ascension & $5^\mathrm{h}15^\mathrm{m}07.6^\mathrm{s}$ \\
        Declination & $-69.37^\circ$ \\
        Coalescence time & GPS $1000$ \\
        Detectors & H1, L1 \\
        Minimum frequency & $20$\,Hz \\
        \hline
    \end{tabular}
\end{table}

We can quantify how informative the likelihood is by the information gained about the distance, the \ac{KL} divergence from the prior $Q$ to the posterior $P$,
\begin{equation}\label{eq:kl-divergence}
    D_{\mathrm{KL}}(P\,\|\,Q) = \int P(\dl)\,\log_2\frac{P(\dl)}{Q(\dl)}\,\mathrm{d}\dl,
\end{equation}
which we give in bits, so that a value of one bit means that the data halve the volume of the prior which the posterior occupies.
A divergence close to zero implies that the gain is negligible.
We evaluate it for the luminosity distance from histograms with 60 bins which each contain an equal fraction of the prior, so that the divergence is $\log_2 60$ minus the entropy of the posterior histogram and is at most $5.9$\,bits.
Bins which are uniform in the logarithm of the distance leave some bins empty in the $\dl^2$ prior, for which the divergence is infinite.
The value depends on the prior bounds as well as on the data, and comparing two halves of the prior samples gives an estimator bias of $0.03$\,bits.

\begin{figure}
    \centering
    {\includegraphics[width=\linewidth]{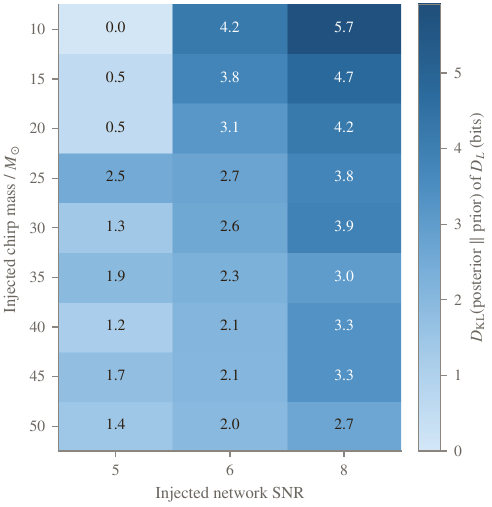}}
    \caption{The information gained about the luminosity distance, the \ac{KL} divergence (in bits) from the prior to the posterior, for each of the simulated injections, as a function of the injected network \ac{SNR} and chirp mass. Each cell is a single noise realisation.}
    \label{fig:js-divergence}
\end{figure}

The gain is close to zero only for the quietest, lowest-mass injection, for which the posterior is almost the prior (figure~\ref{fig:js-divergence}).
At $\rho=6$ it is at least $2.0$\,bits and at $\rho=8$ at least $2.7$\,bits for every chirp mass.
The divergence does not directly indicate how much of the posterior lies where the likelihood cannot constrain the distance.
In figure~\ref{fig:sea-fraction} we show the fraction of posterior samples with a matched-filter \ac{SNR} below $\sqrt{8}$.
It is substantial, between $0.5$ and $0.97$, only for injections with a network \ac{SNR} of five and chirp masses of $20\,M_\odot$ or below, and is small elsewhere, at most $0.12$ for the remaining injections at \ac{SNR} five and at most $0.04$ at \ac{SNR} six.
The fractions for the real events in figure~\ref{fig:sqrt8-events} are comparable.
Since each cell is a single noise realisation, and the prior on the chirp mass is restricted to a range around the injected value, the variation between neighbouring cells should not be over-interpreted.

\subsection{P-P plot for different priors}\label{sec:pp}

A posterior is calibrated for a population of signals only if the prior used in the analysis matches the distribution from which the signals are drawn.
For the same reason a P-P test in which the injection and analysis priors coincide cannot expose a mismatch with the true population~\citep{2024ApJS..273...23V}, and default priors fail the test for simulated astrophysical populations~\citep{2026PhRvD.114b1302M}.

In order to test this we injected $1024$ signals with properties drawn from the prior, using the uniform in comoving volume prior for the luminosity distance, with bounds of $100$ and $8\,000$\,Mpc.
We then analysed these injections using the conditional Jeffreys prior and the marginalised likelihood of equation~\ref{eq:LJ} over the same range.

\begin{figure*}
    \centering
    \includegraphics[width=0.67\textwidth]{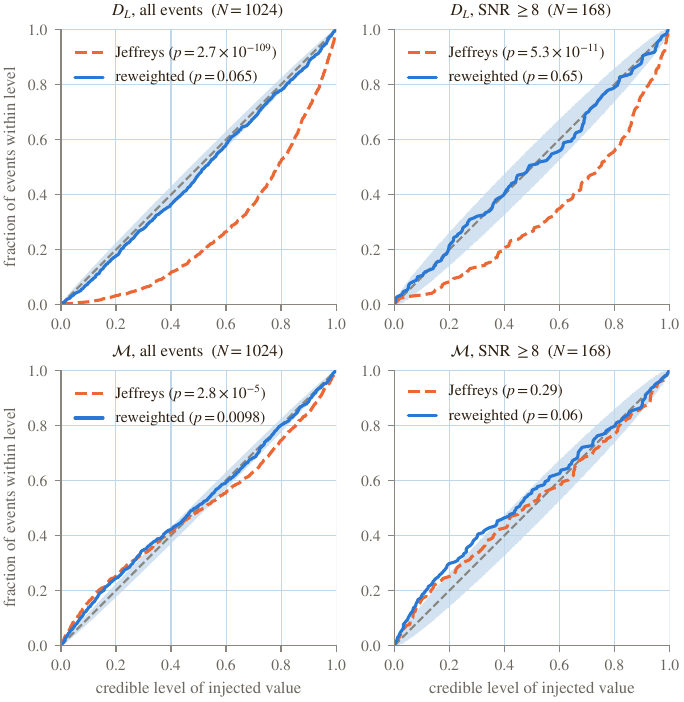}
    \caption{P-P plots for the injected luminosity distance (top) and chirp mass (bottom), for all $1024$ injections (left) and those with a network \ac{SNR} of at least eight (right). The posterior sampled with the Jeffreys distance prior (dashed) is compared with the posterior reweighted to the astrophysical prior (solid); the signals were injected from the astrophysical prior. The shaded band is the $95\%$ binomial interval, and the legends give the Kolmogorov--Smirnov $p$-values.}
    \label{fig:pp}
\end{figure*}

For each signal we computed the credible level of the injected value in the Jeffreys posterior, and in the posterior reweighted to the astrophysical prior as described in section~\ref{sec:reweighting}.
We then show the resulting P-P plots from the entire set of injections in figure~\ref{fig:pp}.
Since the injected distances are drawn from the prior rather than chosen to be detectable, most of the signals are quiet: $550$ of the $1024$ have a network \ac{SNR} below four, and $168$ are above eight.

Given the marked difference between the Jeffreys and the uniform in comoving volume priors, it is 
unsurprising that we find the conditional Jeffreys prior to be strongly miscalibrated when used to analyse these signals.
However, when we reweight the posteriors to the correct prior we find that the measurement of the luminosity distance is unbiased (with Kolmogorov--Smirnov statistic $p=0.065$).

The bias introduced by the conditional Jeffreys prior on the chirp mass is much smaller, but reweighting does not fully resolve this, with the effects more pronounced in lower-mass signals.

\section{Examples from real candidates}\label{sec:examples}

The presence of prior-dominated results in analyses presented in works such as \ac{GWTC5}~\citep{catalog-gwtc-5} is minimal thanks to the choice of threshold used in these works, and the use of the uniform-in-comoving-volume prior for the luminosity distance.
These choices together have eliminated the presence of such results from the catalogues published in the \ac{O4} era.
However, this problem does continue to affect analyses of candidates which fail to cross these significance thresholds.
We demonstrate this with two sets of examples, each from \ac{O3}, which has been extensively analysed since its release in 2020.
The first of these is a subthreshold candidate, S190531n, which is consistent with containing a neutron star as at least one of the binary components.
The second are a handful of \ac{BBH} events which were identified by search pipelines outwith the \ac{LVK}, and which were presented in~\citet{catalog-beyond}.

In both cases the posteriors show signs of prior dominance, and we demonstrate the use of the conditional Jeffreys prior, either through direct sampling, or by reweighting and importance sampling the prior-dominated posteriors.
We then show that we can produce interpretable results for these candidates using the conditional Jeffreys prior.

\subsection{GW190531}\label{sec:190531}

\begin{figure*}

    \includegraphics{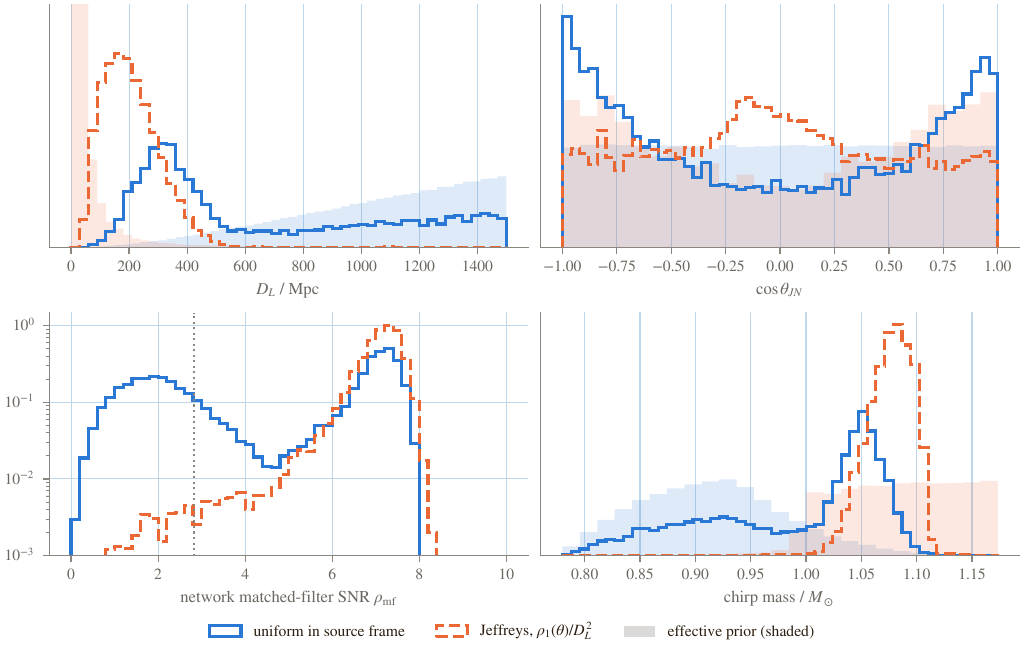}
    \caption{The posterior probability distributions from two analyses of GW190531. 
    The analysis in blue uses the uniform in comoving volume prior, which is the same setup used for analyses in catalogues such as \ac{GWTC4}, and to which results were reweighted prior to release for \ac{GWTC3}.
    The analysis in orange instead uses the conditional Jeffreys prior we introduce in this work.
    We show the marginal posteriors on luminosity distance, the cosine of the inclination angle, and the source-frame chirp mass, alongside the prior distribution for each analysis (shaded regions).
    We also show the effect of each prior on the network matched filter \ac{SNR} distributions of the samples.
    As expected, we find that the conditional Jeffreys prior eliminates the long tail at high distance, and low \ac{SNR}. This makes a significant different to the inferred chirp mass, eliminating the regions $<1M_\odot$.
    \label{fig:s190531n}}

\end{figure*}

GW190531 is a subthreshold trigger from \ac{O3} which was identified as being consistent with being either an \ac{NSBH} or a \ac{BNS} signal in search pipelines.
The signal has low significance, with a \ac{FAR} of $0.41\,{\rm yr}^{-1}$ and a $\pastro=0.28$ in the GstLAL pipeline~\citep{catalog-gwtc2d1,software-gstlal}, although other searches found it with lower significance.
It is also quiet, with a network matched filter \ac{SNR} of $\rho = 10$ in the search.

We analysed this candidate using a similar analysis configuration as used for \ac{GWTC3}.
We used the \texttt{IMRPhenomPv2\_NRTidalv2}~\citep{waveform-imrphenompv2-nrtidal} waveform model with an ROQ likelihood~\citep{PhysRevLett.114.071104,PhysRevD.94.044031,waveform-imrphenompv2-nrtidal-roq}, the uniform in comoving volume prior for the luminosity distance, between $10$ and $1500$\,Mpc, rather than the Euclidean prior used in \ac{GWTC3}, and a low-spin configuration, with the component spin magnitudes restricted to $\chi<0.4$. 
We used bilby~\citep{code-bilby,code-bilby-pipe} and the dynesty nested sampling algorithm~\citep{speagle_dynesty_2020} to perform the inference.
We placed a narrow prior on the chirp mass, which is uniform in component masses, over the range $\mathcal{M}\in (1.0, 1.2)\, M_\odot$.
We also restricted the prior on the tidal parameters to be zero, effectively reproducing the \texttt{IMRPhenomPv2}~\citep{PhysRevLett.113.151101,2019PhRvD.100b4059K} waveform model.
The marginal posterior distributions on $\dl$, the cosine of the inclination angle, and the chirp mass are shown in blue in figure~\ref{fig:s190531n}.

The $\dl$ posterior clearly shows the effects of prior dominance, with a long tail extending to very high distances: at $1500$\,Mpc the optimal \ac{SNR} for signals allowed by the chirp mass prior would be around $\rho=0.6$.
We find that $38\%$ of the posterior samples have $\rhomf<\sqrt{8}$, indicating that a significant fraction of the posterior is prior dominated.

We also analysed this candidate using the same analysis configuration, but using the conditional Jeffreys prior for the luminosity distance.
The marginal posterior distributions from this analysis are shown in orange in figure~\ref{fig:s190531n}.
The use of an ROQ likelihood required us to perform this in two steps.
First we performed the analysis using a prior with the form $\dl^{-2}$, and then we reweighted the samples to the conditional Jeffreys prior by applying the weighting $\rho_1(\theta) = \rho_{\rm opt} \dl$ to each sample.
The effective sample size of the weighted samples is $11{,}096$, which is $82\%$ of the original $13{,}591$ samples.
In contrast to the uniform in comoving volume prior, the posteriors from this analysis produce a finite mode, and a result for which statistics such as the median and credible intervals are interpretable.

Using the conditional Jeffreys prior, we find this event has a median luminosity distance of $196^{+163}_{-140}$\,Mpc (median and $90\%$ highest probability density (HPD) interval), compared to $467^{+933}_{-325}$\,Mpc when using the uniform in comoving volume prior. Eliminating the low SNR tail significantly changes the astrophysical interpretation: the inferred chirp mass of the system becomes $1.077^{+0.031}_{-0.035}\,M_\odot$ in the source frame, compared to $1.025^{+0.063}_{-0.170}\,M_\odot$ with the uniform in comoving volume prior, which is more in line with the expectation of components $>1M_\odot$.

\begin{figure*}
    \centering
    \includegraphics[width=\textwidth]{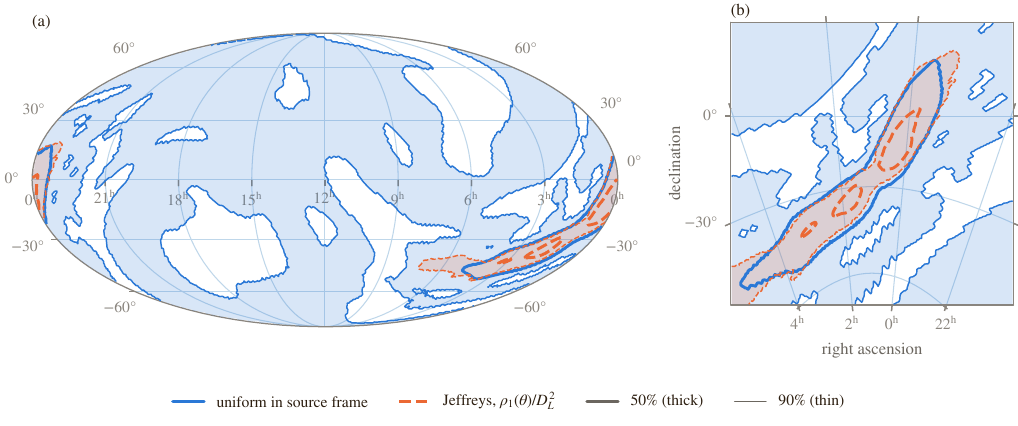}
    \caption{The sky localisation of GW190531 from the analyses of figure~\ref{fig:s190531n}, using the uniform in comoving volume prior (blue, solid) and the conditional Jeffreys prior (orange, dashed).
    The thick and thin contours enclose the $50\%$ and $90\%$ credible regions.
    (a) The whole sky, in equatorial coordinates.
    (b) The region around the arc which contains the Jeffreys $90\%$ region.}
    \label{fig:s190531n-skymap}
\end{figure*}

The change of prior also has a large effect on the sky localisation of the candidate.
In figure~\ref{fig:s190531n-skymap} we show a comparison of the skymaps under each prior, which we estimate from $4000$ equally weighted samples from each analysis using the clustered kernel density estimate in \texttt{ligo.skymap}~\citep{PhysRevD.93.024013}.
The $90\%$ credible area is $29{,}000\,\mathrm{deg}^2$, or $70\%$ of the sky, with the uniform in comoving volume prior, and $1400\,\mathrm{deg}^2$ with the conditional Jeffreys prior, a reduction by a factor of $20$; the $50\%$ areas are $1300$ and $270\,\mathrm{deg}^2$.
The two maps have their most confident region along the same arc, which contains the entire conditional Jeffreys $90\%$ region.
Much of the difference comes from the low-\ac{SNR} samples, for which the data barely constrain the arrival times at the detectors, so that they are spread across the sky: simply discarding the samples with $\rhomf<\sqrt{8}$ from the uniform in comoving volume analysis reduces its $90\%$ area to $7700\,\mathrm{deg}^2$.
The skymap created from the conditional Jeffreys prior shows where the data place a signal of appreciable strength which is the more useful quantity for planning follow-up observations.

We will discuss this candidate, alongside other subthreshold candidates from \ac{O3} in a future work~\claude{\citepalias{wwwb-subthreshold}}, however, we show here that the use of the conditional Jeffreys prior can be used to produce interpretable results for it.

\subsection{O3 near-threshold \acp{BBH}}\label{sec:beyond-events}

\begin{figure*}
    \centering
    \includegraphics[width=\linewidth]{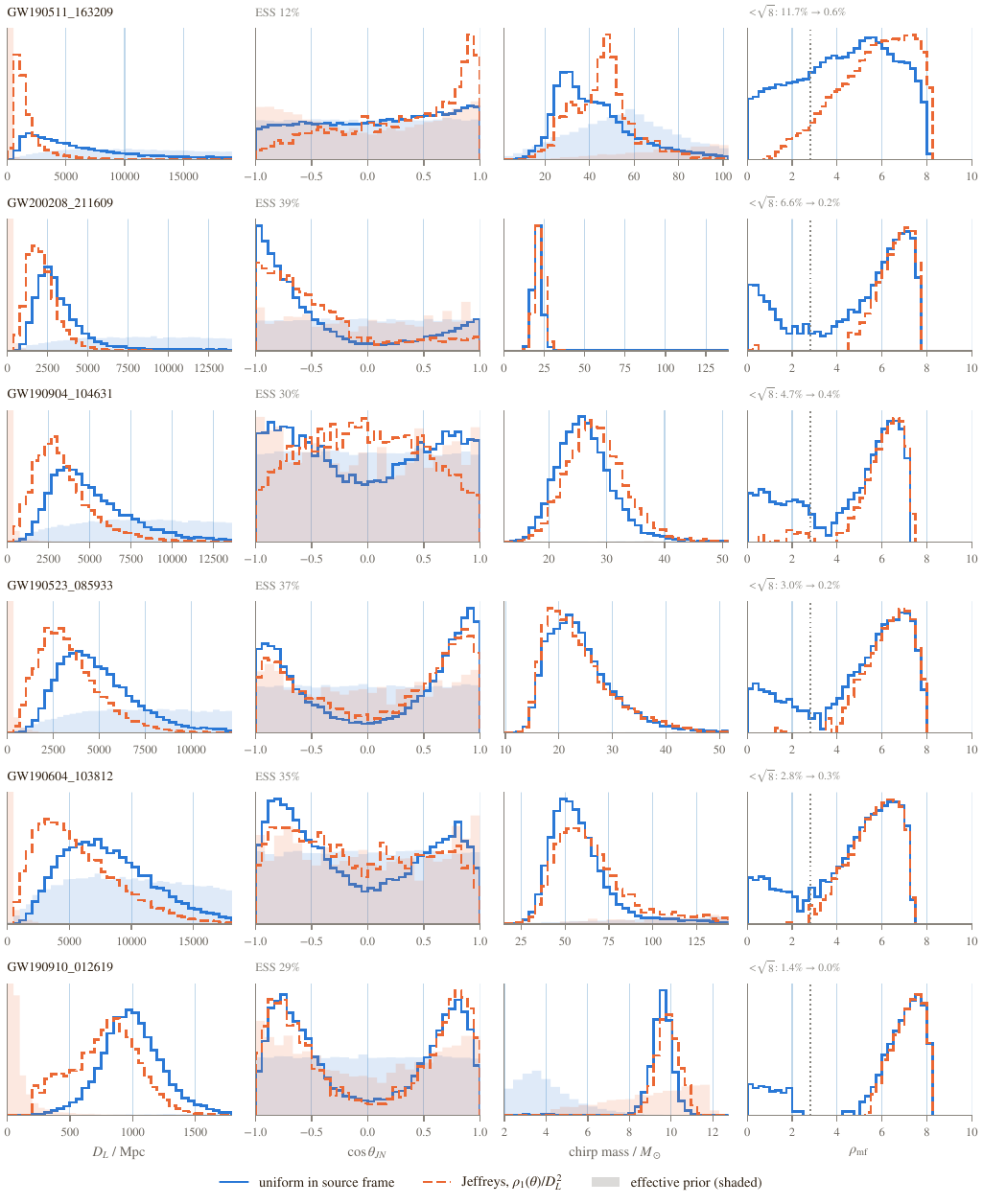}
    \caption{
        The marginal posterior distributions on $\dl$, $\cos \theta_{JN}$, and chirp mass for the six \ac{BBH} candidates identified from~\citet{catalog-beyond} which have the largest percentage of samples with $\rhomf < \sqrt{8}$.
        In blue we show the posteriors from the published analyses from \citet{catalog-beyond}, which used a prior uniform in comoving volume and source-frame time, and in orange we show the same samples reweighted to the conditional Jeffreys prior.
        For each analysis we show the prior distribution as a shaded region.
    \label{fig:beyond-events-posteriors}}
\end{figure*}

A number of signal candidates during \ac{O3} were identified by searches which were independent of the \ac{LVK}.
These are summarised in \citet{catalog-beyond}, which also reanalyses the triggers using analysis configurations comparable to \ac{GWTC3}, except that the uniform-in-comoving volume prior was used for sampling in these analyses rather than the Euclidean prior used in \ac{GWTC3}.

While the $\rhomf = \sqrt{8}$ threshold does not strictly apply to analyses using the uniform in comoving volume prior, it nevertheless provides a simple diagnostic to initially identify posteriors which may involve prior dominance.
We checked samples from the posteriors for all events provided in \citet{catalog-beyond}, and identified the six events which had the largest percentage of samples with $\rhomf < \sqrt{8}$.

We show (in blue) the marginal posteriors on $\dl$, $\cos \theta_{JN}$, and chirp mass in figure~\ref{fig:beyond-events-posteriors}.
The consequences of the prior dominance are most obvious in the distance posterior for GW190511\_163209, which has a clear long-distance tail.
The effects are more subtle for other candidates.

Rather than performing new analyses for each of these candidates, we reweighted the published posterior samples to the conditional Jeffreys prior, and plot the results in orange on figure~\ref{fig:beyond-events-posteriors}.
The most obvious effect of this is in the marginal posteriors on $\dl$, especially for GW190511\_163209, which goes from having a distance of {$4800^{+8910}_{-4440}$\,Mpc (median and $90\%$ HPD interval) in the published posterior to $1310^{+4340}_{-950}$\,Mpc} in the reweighted one.
In table~\ref{tab:beyond-events} we show the published and reweighted posterior values for each quantity, alongside the percentage of samples where $\rho_{\rm mf} < \sqrt{8}$.

\begin{table*}
\caption{The median and $90\%$ highest probability density (HPD) interval on the luminosity distance, the source-frame chirp mass, and $\cos\theta_{JN}$ for the events in figure~\ref{fig:beyond-events-posteriors}. For each quantity the left column is the posterior published in~\citet{catalog-beyond} sampled with a prior uniform in comoving volume and time (uniform in source frame; USF), and the right column is the same samples reweighted to the conditional Jeffreys prior. The final columns give the percentage of samples with $\rho_{\rm mf}<\sqrt{8}$ in the same two cases, and the effective sample size resulting from reweighting the published samples to the conditional Jeffreys prior, as a percentage of the published samples.}
\label{tab:beyond-events}
\centering
\small
\setlength{\tabcolsep}{4pt}
\begin{tabular}{lcccccccccc}
\hline
 & \multicolumn{2}{c}{$\dl$ /Mpc} & \multicolumn{2}{c}{$\Mc$ /$M_\odot$} & \multicolumn{2}{c}{$\cos\theta_{JN}$} & \multicolumn{2}{c}{$\rhomf<\sqrt{8}$ /\%} & ESS \\
Event & USF & Jeffreys & USF & Jeffreys & USF & Jeffreys & USF & Jeffreys & /\% \\
\hline
GW190511\_163209 & $4800^{+8910}_{-4440}$ & $1310^{+4340}_{-950}$ & $39.2^{+35.5}_{-23.6}$ & $46.3^{+22.9}_{-25.1}$ & $0.12^{+0.85}_{-0.96}$ & $0.43^{+0.57}_{-1.21}$ & $11.7$ & $0.6$ & $12$ \\
GW200208\_211609 & $2910^{+3360}_{-2230}$ & $2030^{+1570}_{-1390}$ & $21.0^{+35.6}_{-14.5}$ & $22.0^{+6.1}_{-5.5}$ & $-0.67^{+1.64}_{-0.33}$ & $-0.59^{+1.35}_{-0.41}$ & $6.6$ & $0.2$ & $39$ \\
GW190904\_104631 & $4580^{+4620}_{-3210}$ & $3010^{+3070}_{-2440}$ & $26.1^{+15.8}_{-15.3}$ & $28.0^{+10.1}_{-9.5}$ & $-0.03^{+0.95}_{-0.91}$ & $-0.04^{+0.85}_{-0.84}$ & $4.7$ & $0.4$ & $30$ \\
GW190523\_085933 & $4580^{+3870}_{-3050}$ & $3120^{+2920}_{-2390}$ & $23.2^{+18.0}_{-12.0}$ & $22.5^{+13.0}_{-10.0}$ & $0.40^{+0.60}_{-1.38}$ & $0.33^{+0.67}_{-1.27}$ & $3.0$ & $0.2$ & $37$ \\
GW190604\_103812 & $7730^{+6320}_{-5470}$ & $5060^{+5970}_{-4320}$ & $55.4^{+31.5}_{-25.5}$ & $61.4^{+43.1}_{-30.2}$ & $-0.20^{+1.09}_{-0.75}$ & $-0.13^{+0.97}_{-0.81}$ & $2.8$ & $0.3$ & $35$ \\
GW190910\_012619 & $990^{+550}_{-530}$ & $800^{+390}_{-510}$ & $9.6^{+0.8}_{-0.9}$ & $9.8^{+1.0}_{-0.9}$ & $-0.14^{+1.08}_{-0.79}$ & $-0.07^{+1.06}_{-0.90}$ & $1.4$ & $0.0$ & $29$ \\
\hline
\end{tabular}
\end{table*}

For these events, under the prior used for the published analyses between $1.4\%$ and $11.7\%$ of the samples have $\rhomf<\sqrt{8}$; after reweighting to the conditional Jeffreys prior none of them has more than $0.6\%$.
The median distance falls by between $19\%$ and $73\%$, and by roughly a third for four of the six.
For GW190511\_163209 the upper end of the $90\%$ HPD interval on the distance falls from $13.7$ to $5.7$\,Gpc, while the lower end is unchanged at $0.36$\,Gpc.
The upper end of the published interval largely reflects the prior at large distance, rather than the data.
The source-frame masses are obtained from the detector-frame masses but depend on the luminosity distance.
Therefore these also change, by up to $7\,M_\odot$.
The high-mass tail of GW200208\_211609 disappears as a result of the reweighting.
The presence of prior-dominated samples in the published posteriors for these events should be considered when interpreting these signals.
GW190511\_163209 was identified as having a large difference between the results presented in~\citet{catalog-beyond} and \citet{olsen_new_2022} where it was originally identified and analysed.
We believe that this justifies further examination of low-significance  candidates such as these with analysis under the conditional Jeffreys prior.

\section{Conclusions}\label{sec:conclusion}

We have considered \ac{PE} of \ac{GW} signals which are quiet enough that the posterior on the luminosity distance is dominated by information from the prior.
For a prior $\propto\dl^2$, with other parameters fixed we have shown, in section~\ref{sec:problem} that the posterior has a finite mode only if the matched-filter \ac{SNR} is at least $\sqrt{8}$.
Below this critical \ac{SNR} the posterior will contain a long tail of samples which are derived from the prior, making quantities such as the median unhelpful as a summary statistic.
For two published \ac{GW} candidates from \ac{GWTC3}, GW200308\_173609 and GW200322\_091133, a large fraction of the posterior samples from the main analysis lie below this threshold ($36\%$ and $60\%$ respectively),
and so summary statistics from their posteriors reflect the prior bounds as much as the data.
For these two events reweighting from the Euclidean distance prior used when performing inference to one which was uniform in comoving volume resolves this problem, and both events' posteriors have finite modes.
In section~\ref{sec:cosmo-threshold} we showed why this is, as the critical \ac{SNR} for this prior is lower.

However, this change of prior does not solve the problem altogether, it simply changes the threshold at which the problem will arise.
To address this we introduced the conditional Jeffreys prior on the luminosity distance in section~\ref{sec:jeffreys}.
This prior is immune to this problem by being minimally informative, and so the shape of the marginal posterior is informed only by the likelihood, and thus the data. 
The conditional prior on the distance is the reference prior: for fixed values of the other parameters, and when the range of optimal SNR is wide compared with the noise, it maximises the expected information gained about the distance from the data (see section~\ref{sec:jeffreys}).
This prior also has the attractive quality of allowing the luminosity distance parameter to be analytically marginalised and later reconstructed. While the conditional Jeffreys prior is well-motivated from the perspective of information theory it is clearly incorrect astrophysically, as it ignores our knowledge of the geometry of the Universe.

Any statistic generated from a posterior which used this prior must therefore be treated separately from those generated under an astrophysically-motivated prior.

We compared the conditional Jeffreys prior with a similar concept which has been considered in the field in the past, the so-called Malmquist prior (introduced in section~\ref{sec:malmquist}).
Our approach avoids objections to this prior based on the ``double-counting'' of the data.

The Jeffreys prior can also be extended to cover other parameters of the signal.
While for many of these it would be excessively uninformative, an obvious extension to our work would be to perform the analysis with a joint Jeffreys prior in the luminosity distance and the source inclination angle, as these parameters are strongly related through their effect on the amplitude of the signal.

Which prior is appropriate depends upon the question being asked of the posterior.
For an astrophysical estimate, such as the redshift or source-frame masses of an event, an astrophysical prior is required, and the best available is the population itself, marginalised over its hyperparameters; the priors of section~\ref{sec:distance-priors} are fixed-hyperparameter approximations to it, each assuming a particular redshift evolution of the rate. Candidates below the detection threshold may have no fitted population to condition on, and the same is true of \ac{BNS} signals, for which the observed population is small and the population-informed approach is correspondingly less well constrained.
For other uses a reference prior is a better choice. 
For example, a follow-up observation is concerned with where the data, and not the prior, places probability. 
For example, in the localisation of a faint signal the long high-distance tail is prior-dominated and of limited interest to a follow-up observer.
We showed this with the localisation of a sub-threshold \ac{BNS} signal in section~\ref{sec:190531}, and the dramatic difference it makes to the skymap in figure~\ref{fig:s190531n-skymap}.
The conditional Jeffreys result is not a substitute for an astrophysical posterior; it is the likelihood-dominated quantity from which one can be recovered once the population is known.

The landscape of \ac{GW} analysis is becoming increasingly sophisticated.
Search algorithms are becoming more capable of identifying signal candidates below traditional detection thresholds.
Techniques to perform signal inference in the presence of non-Gaussian noise artefacts in the data are also maturing~\citep{glitchflow,gbilby}, as are approaches to performing population inference by treating a population as a mixture of real signals and noise artefacts~\citep{gaebelDiggingPopulationCompact2019a, 2023MNRAS.523.5972H}.
Together these innovations are making analysis of subthreshold candidates a viable area of research.

Recent catalogues have shown that near-threshold events are often difficult to interpret when the astrophysical prior on distance is used. 
Our work introduces a principled alternative prior that does not double-count the data, but does isolate the mode in parameter space that corresponds to the detected trigger.

\section*{Data availability}

Data used in this paper are available to download from Zenodo~\citep{williams_2026_23807340}.

\section*{Acknowledgments}

This research has made use of data or software obtained from the Gravitational Wave Open Science Center (gwosc.org), a service of the LIGO Scientific Collaboration, the Virgo Collaboration, and KAGRA. This material is based upon work supported by NSF's LIGO Laboratory which is a major facility fully funded by the National Science Foundation, as well as the Science and Technology Facilities Council (STFC) of the United Kingdom, the Max-Planck-Society (MPS), and the State of Niedersachsen/Germany for support of the construction of Advanced LIGO and construction and operation of the GEO600 detector. Additional support for Advanced LIGO was provided by the Australian Research Council. Virgo is funded, through the European Gravitational Observatory (EGO), by the French Centre National de Recherche Scientifique (CNRS), the Italian Istituto Nazionale di Fisica Nucleare (INFN) and the Dutch Nikhef, with contributions by institutions from Belgium, Germany, Greece, Hungary, Ireland, Japan, Monaco, Poland, Portugal, Spain. KAGRA is supported by Ministry of Education, Culture, Sports, Science and Technology (MEXT), Japan Society for the Promotion of Science (JSPS) in Japan; National Research Foundation (NRF) and Ministry of Science and ICT (MSIT) in Korea; Academia Sinica (AS) and National Science and Technology Council (NSTC) in Taiwan.

DW and JV are supported under Science and Technology Facilities Council (STFC) grant UKRI2487.

The authors acknowledge the use of the LLM tool Claude during preparation of this manuscript, however any outputs from the model have been edited appropriately by the authors.

The authors also thank Michael~J. Williams for useful discussions on reweighting nested samples between priors; we also wish to thank both Michael and Matthew Mould for comments on the manuscript and suggested references.

The analyses presented in this paper were run on the computing resources at the Institute for Gravitational Research at the University of Glasgow. 
We acknowledge the following software packages which were integral to this analysis: asimov~\citep{code-asimov}, bilby~\citep{code-bilby}, bilby\_pipe~\citep{code-bilby-pipe}, bayeswave~\citep{cornish_bayeswave_2015}, pesummary~\citep{code-pesummary}, dynesty~\citep{speagle_dynesty_2020}, numpy~\citep{software-numpy}, scipy~\citep{scipy}, matplotlib~\citep{software-matplotlib}, htcondor~\citep{software-htcondor}, and astropy~\citep{software-astropy-1,software-astropy-2,software-astropy-3}.

This document has been assigned LIGO Document Control Centre identifier LIGO-P2600602.

\bibliography{references}
\bibliographystyle{mnras}
\end{document}